\documentstyle[prd,aps]{revtex}

\makeatletter                    
\@addtoreset{equation}{section}  
\makeatother

\begin{document}
\title{Stochastic dynamics of correlations in quantum field theory: From
Schwinger-Dyson to Boltzmann-Langevin equation}
\author{Esteban Calzetta}
\address{{\small FCEN y IAFE, Universidad de Buenos Aires, Argentina}\\
B. L. Hu\\
{\small Department of Physics, University of Maryland, College Park,
Maryland 20742}}
\date{{\small umdpp 98-121, Mar. 5, 1999}}
\maketitle

\begin{abstract}
The aim of this paper is two-fold: in probing the statistical mechanical
properties of interacting quantum fields, and in providing a field
theoretical justification for a stochastic source term in the Boltzmann
equation. We start with the formulation of quantum field theory in terms of
the Schwinger - Dyson equations for the correlation functions, which we
describe by a closed-time-path master ($n = \infty PI$) effective action.
When the hierarchy is truncated, one obtains the ordinary closed-system of
correlation functions up to a certain order, and from the nPI effective
action, a set of time-reversal invariant equations of motion. But when the
effect of the higher order correlation functions is included (through e.g.,
causal factorization-- molecular chaos -- conditions, which we call
'slaving'), in the form of a correlation noise, the dynamics of the lower
order correlations shows dissipative features, as familiar in the
field-theory version of Boltzmann equation. We show that
fluctuation-dissipation relations exist for such effectively open systems,
and use them to show that such a stochastic term, which explicitly
introduces quantum fluctuations on the lower order correlation functions,
necessarily accompanies the dissipative term, thus leading to a
Boltzmann-Langevin equation which depicts both the dissipative and
stochastic dynamics of correlation functions in quantum field theory.
\end{abstract}


\section{Introduction}

The main result of this paper is a derivation of the Boltzmann - Langevin
equation as the correct description of the kinetic limit of quantum field
theory. We use for this derivation a generalized formulation of quantum
field theory in terms of the Schwinger - Dyson equations but supplemented
with stochastic terms which explicitly introduce the effects of quantum
fluctuations on low order correlation functions.

The significance of such an inquiry is two-fold: In probing the statistical
mechanical properties of interacting quantum fields, and in providing a
field theoretical justification for a stochastic source term in the
Boltzmann equation. The former has been investigated mainly for quantum
mechanical, not field-theoretical, systems \cite{Balescu,Akhiezer} (see
however, \cite{Yaffe,Haag}) and the latter primarily for a classical gas 
\cite{KacLog,Spohn}. Extending previous studies to quantum fields is
essential in the establishment of a quantum field theory of nonequilibrium
processes. Previous work on this subject \cite
{kadbaym,dubois,limcl,CH88,DMHW,mrow,hen,bll} showed how the Boltzmann
equation can be derived from first-principles in quantum field theory and
investigated its dissipative properties. In this paper we focus on the
fluctuation and noise aspects in the derivation of a stochastic Boltzmann
equation from quantum field theory \cite{abe,reinhard,greinleu}.\\ 

\noindent{\it Fluctuations in Composite Operators}\\

There are a variety of problems in nonequilibrium field theory which are
most naturally described in terms of the unfolding of composite operators,
the familiar lowest order ones being the particle number and the energy
momentum densities and their fluctuations. The usual approach to these
problems assumes that these operators have small fluctuations around their
expectation values, in which case they can be expressed in terms of the
Green functions of the theory. However, when fluctuations are large,
typically when corrrelations among several particles are important, this
approximation breaks down.

A familiar example is critical phenomena: by choosing a suitable order
parameter to describe the different phases, one can obtain a wealth of
information on the phase diagram of a system. But to study the dynamics of a
phase transition, especially in the regime where fluctuations get large, the
single order parameter must be replaced by a locally defined field obeying a
stochastic equation of motion, for example, a time - dependent Ginzburg -
Landau equation with noise (which though often put in by hand, should in
theory be derived from fluctuation - dissipation considerations). The same
phenomenon occurs more generally in effective field theories, where the
light fields are randomized by the back reaction from the heavy fields \cite
{CH97}, and in semiclassical theories, where the classical field (for
example, the gravitational field in the Early Universe) is subject to random
driving forces from activities in the quantum field, such as particle
creation \cite{CH94}.

The influence of noise on the classical dynamics of a quantum system is
discussed at length by Gell-Mann and Hartle \cite{GelHar2}; the conversion
of quantum fluctuations to classical noise is discussed by a number of
authors \cite{Belgium,Banff,CH95,Kiefer,Halliwell}. This scheme was also
used by us for the study of decoherence of correlation histories and
correlation noise in interacting field theories \cite{dch,cddn} and related
work by others using projection operators \cite{Anas}.

Therefore in the more general context illustrated by these examples, we
shall explore models where the dynamics is described not in terms of the
actual Green functions $G^{ab}$ (we use closed - time - path (CTP)
techniques and notation, described more fully in the Appendix\cite{ctp}) but
a stochastic kernel ${\bf G}^{ab}$, whose expectation value reduces to $%
G^{ab},$ while its fluctuations reproduce the quantum fluctuations in the
binary products of field operators.

More concretely, consider a theory of a scalar field $\phi ^a$ (we use a
condensed notation where the index $a$ denotes both a space - time point and
one or the other branch of the time path -- see Appendix). The CTP action is 
$S=S\left[ \phi ^1\right] -S^{*}\left[ \phi ^2\right] $. Introduce the
generating functional

\begin{equation}
Z\left[ K_{ab}\right] =e^{iW\left[ K_{ab}\right] }=\int D\phi
^a\;e^{i\left\{ S+\frac 12K_{ab}\phi ^a\phi ^b\right\} }  \label{genfun}
\end{equation}
then

\begin{equation}
G^{ab}=\left\langle \phi ^a\phi ^b\right\rangle =2\left. \frac{\delta W}{%
\delta K_{ab}}\right| _{K=0}  \label{greenfun}
\end{equation}
but also

\begin{equation}
\left. \frac{\delta ^2W}{\delta K_{ab}\delta K_{cd}}\right| _{K=0}=\frac i4%
\left\{ \left\langle \phi ^a\phi ^b\phi ^c\phi ^d\right\rangle -\left\langle
\phi ^a\phi ^b\right\rangle \left\langle \phi ^c\phi ^d\right\rangle \right\}
\label{fluc}
\end{equation}
This suggests viewing the stochastic kernel ${\bf G}^{ab}$ as a Gaussian
process defined (formally) by the relationships

\begin{equation}
\left\langle {\bf G}^{ab}\right\rangle =\left\langle \phi ^a\phi
^b\right\rangle ;\qquad \left\langle {\bf G}^{ab}{\bf G}^{cd}\right\rangle
=\left\langle \phi ^a\phi ^b\phi ^c\phi ^d\right\rangle  \label{stocg}
\end{equation}
Or else, calling

\begin{equation}
{\bf G}^{ab}=G^{ab}+\Delta ^{ab}  \label{delta}
\end{equation}

\begin{equation}
\left\langle \Delta ^{ab}\right\rangle =0;\qquad \left\langle \Delta
^{ab}\Delta ^{cd}\right\rangle =-4i\left. \frac{\delta ^2W}{\delta
K_{ab}\delta K_{cd}}\right| _{K=0}  \label{delta2}
\end{equation}

To turn the intuitive ansatz Eqs. (\ref{stocg}) and (\ref{delta2}) into a
rigorous formalism we must deal with the obvious fact that we are
manipulating complex expressions; in particular, it is not clear the $\Delta$%
s define a stochastic process at all. However, for our present purposes it
will prove enough to deal with the propagators {\it as if } they were real
quantities. The reason is that we are primarily concerned with the large
occupation numbers or semiclassical limit, where the propagators do become
real. We will see that this prescription will be sufficient to extract
unambiguous results from the formal manipulations below.\\

\noindent{\it Stochastic Boltzmann Equation}\\

We can define the process $\Delta ^{ab}$ also in terms of a stochastic
equation of motion. Consider the Legendre transform of $W$, the so - called
2 particle irreducible (2PI) effective action (EA)

\begin{equation}
\Gamma \left[ G^{ab}\right] =W\left[ K_{ab}^{*}\right] -\frac 12K_{ab}^{*}%
{\bf G}^{ab};\qquad K_{ab}^{*}=-2\frac{\delta \Gamma }{\delta {\bf G}^{ab}}
\label{tpiea}
\end{equation}
We have the identities

\begin{equation}
\frac{\delta \Gamma }{\delta G^{ab}}=0;\qquad \frac{\delta ^2W}{\delta
K_{ab}\delta K_{cd}}=\frac{-1}4\left[ \frac{\delta ^2\Gamma }{\delta {\bf G}%
^{ab}\delta {\bf G}^{cd}}\right] ^{-1}  \label{ident}
\end{equation}
the first of which is just the Schwinger - Dyson equation for the
propagators; we therefore propose the following equations of motion for $%
{\bf G}^{ab}$

\begin{equation}
\frac{\delta \Gamma }{\delta {\bf G}^{ab}}=\frac{-1}2\kappa _{ab}
\label{lan2pi}
\end{equation}
where $\kappa _{ab}$ is a stochastic nonlocal Gaussian source defined by

\begin{equation}
\left\langle \kappa _{ab}\right\rangle =0;\qquad \left\langle \kappa
_{ab}\kappa _{cd}\right\rangle =4i\left[ \frac{\delta ^2\Gamma }{\delta
G^{ab}\delta G^{cd}}\right] ^{\dagger }  \label{noisecor}
\end{equation}
If we linearize Eq. (\ref{lan2pi}) around $G$, then the correlation Eq. (\ref
{noisecor}) for $\kappa $ implies Eq. (\ref{delta2}) for $\Delta $.
Consistent with our recipe of handling $G$ as if it were real we should
treat $\kappa $ also as if it were a real source.

It is well known that the noiseless Eq. (\ref{lan2pi}) can be used as a
basis for the derivation of transport equations in the near equilibrium
limit. Indeed, for a $\lambda \phi ^4$ type theory, the resulting equation
is simply the Boltzmann equation for a distribution function $f$ defined
from the Wigner transform of $G^{ab}$ (details are given below). We shall
show in this paper that the full stochastic equation (\ref{lan2pi}) leads,
in the same limit, to a Boltzmann - Langevin equation, thus providing the
microscopic basis for this equation in a manifestly relativistic quantum
field theory.

Let us first examine some consequences of Eq. (\ref{noisecor}). For a free
field theory, we can compute the 2PI EA explicitly (derivation in Section V)

\begin{equation}
\Gamma \left[ G^{ab}\right] =\frac{-i}2\ln \left[ Det{\bf G}\right] -\frac 12%
c_{ab}\left( -\Box +m^2\right) {\bf G}^{ab}\left( x,x\right)  \label{free2pi}
\end{equation}
where $c_{ab}$ is the CTP metric tensor (see the Appendix). We immediately
find

\begin{equation}
\frac{\delta ^2\Gamma }{\delta G^{ab}\delta G^{cd}}=\frac i2\left(
G^{-1}\right) _{ac}\left( G^{-1}\right) _{db}  \label{secondvar}
\end{equation}
therefore

\begin{equation}
\left\langle \Delta ^{ab}\Delta ^{cd}\right\rangle =i\left[ \frac{\delta
^2\Gamma }{\delta G^{ab}\delta G^{cd}}\right] ^{-1}=G^{ac}G^{db}+G^{da}G^{bc}
\label{freefluc}
\end{equation}
an eminently sensible result. Observe that the stochastic source does not
vanish in this case, rather

\begin{equation}
\left\langle \kappa _{ab}\kappa _{cd}\right\rangle
=G_{ac}^{-1}G_{db}^{-1}+G_{da}^{-1}G_{bc}^{-1}  \label{freecor}
\end{equation}
However

\begin{equation}
\left( G^{-1}\right) _{ac}\sim -ic_{ac}\left( -\Box +m^2\right)
\label{freeinvprop}
\end{equation}
does vanish on mass - shell. Therefore, when we take the kinetic theory
limit, we shall find that for a free theory, there are no on - shell
fluctuations of the distribution function. 
For an interacting theory this is no longer the case.

The physical reason for this different behavior is that the evolution of the
distribution function for an interacting theory is dissipative, and
therefore basic statistical mechanics considerations call for the presence
of fluctuations\cite{fdt}. Indeed it is this kind of consideration which led
us to think about a Boltzmann - Langevin equation in the first place. This
is fine if one takes a statistical mechanical viewpoint, but one is used to
the idea that quantum field theories are unitary and complete with no
information loss, so how could one see dissipation or noise?

In field theory there is a particular derivation of the self consistent
dynamics for Green functions which resolves this puzzle, namely when the
Dyson equations are derived from the variation of a nonlocal action
functional, the two-particle irreducible effective action (2PI-EA). This was
originally introduced \cite{pi2} as a convenient way to perform
nonperturbative resummation of several Feynman graphs. When cast in the
Schwinger - Keldysh ''closed time path'' (CTP) formulation \cite{ctp}, it
guarantees real and causal evolution equations for the Green functions of
the theory. It is conceptually clear if one begins with a ''master''
effective action (MEA) \cite{cddn} where all Green functions of the theory
appear as arguments, and then systematically eliminate all higher-than-two
point functions to arrive at the 2PIEA.

To us, the correct approach is to view the two point functions as an open
system\cite{qos,qbm,if} truncated from the hierarchy of correlation
functions obeying the set of Schwinger-Dyson equations but interacting with
an environment of higher irreducible correlations, whose averaged effect
brings about dissipation and whose fluctuations give rise to the correlation
noise \cite{cddn}. This is the conceptual basis of our program.\\

\noindent{\it Fluctuations and Dissipation}\\

It has long been known in statistical physics that the equilibrium state is
far from being static; quite the opposite, it is the fluctuations around
equilibrium which underlie and give meaning to such phenomena as Brownian
motion \cite{qbm} and transport processes, and determine the responses (such
as the heat capacity and susceptibility) of the system in equilibrium. The
condition that equilibrium constantly reproduces itself in the course of all
these activities means that the equilibrium state is closely related both to
the structure of the equilibrium fluctuations and to the dynamical processes
by which equilibrium sustains itself; these simple but deep relations are
embodied in the so-called fluctuation - dissipation theorems: If a
fluctuating system is to persist in the neighborhood of a given equilibrium
state, then the overall dissipative processes in the system (due mainly to
its interaction with its environment) are determined. Vice versa, if the
dissipative processes are known, then we may describe the properties of
equilibrium fluctuations without detailed knowledge of the system's
microscopic structure. This is the aspect of the fluctuation - dissipation
relations which guided Einstein in his pioneering analysis of the
corpuscular structure of matter \cite{Einstein}, Nyquist in his stochastic
theory of electric resistivity \cite{Nyquist}, and Landau and Lifshitz to
the theory of hydrodynamical fluctuations \cite{ll57}.

These ideas apply to systems described by a few macroscopic variables, as
well as to systems described by an infinite number of degrees of freedom
such as a few long wavelength modes (as in hydrodynamics) or a single
particle distribution function (as in kinetic theory). In this later case,
the dynamics is described by a dissipative Boltzmann equation, and thereby
we are to expect that there will be nontrivial fluctuations in equilibrium.
The stochastic properties of the Boltzmann equation has been discussed by
Zwanzig, Kac and Logan, and others \cite{KacLog,StochBoltz}.

The Boltzmann equation can be retrieved also as a description of the long
range, near equilibrium dynamics of field theories \cite{CH88}. In this
kinetic theory regime, where there is a clear separation of microscopic and
macroscopic scales the field may be described in terms of quasi-particles,
whose distribution function obeys a Boltzmann equation \cite{CH88,Yaffe}.
Formally, the one particle distribution function is introduced as a partial
Fourier transform of a suitable Green function of the field. The same
arguments which lead to a fluctuating Boltzmann equation in the general
case, lead us to expect fluctuations in this limiting case of field theory
as well.

The end result of our investigation is a highly nonlinear, explicitly
stochastic Dyson equation for the Green functions. By going to the kinetic
theory limit, we derive a stochastic Boltzmann equation, and the resulting
noise may be compared with that required by the fluctuation - dissipation
relation. Here we see clearly this contrast between the predictions of field
theory with and without statistical physics considerations.

In this paper, we shall concentrate on the issue of what kind of
fluctuations may be convincingly derived from the 2PI-CTP-EA for Green
functions, and how they compare to the fluctuation - dissipation noise in
the kinetic theory limit. Given the complexity of the subject, we shall
adopt a line of development which favors at least in the beginning ease of
understanding over completeness. That is, instead of starting from the
master effective action of n point functions and work our way down in a
systematic way, we shall begin with the Boltzmann equation for one -particle
distributions and work our way up.

In the next section, we present briefly the fluctuation - dissipation
theorem in a nonrelativistic context, and use it to derive the fluctuating
Boltzmann equation. The discussion, kept at the classical level, simply
reviews well established results in the theory of the Boltzmann equation.
Section III reviews the basic tenets of nonequilibrium quantum field theory
as it concerns the dynamics of correlations, and the retrieval of the
Boltzmann equation therefrom. We refrain from using functional methods, so
as to keep the discussion as intuitive as possible.

Section IV discusses how the functional derivation of the Schwinger - Dyson
hierarchy suggests that these equations ought to be enlarged to include
stochastic terms. By going through the kinetic limit we use these results to
establish a comparison with the purely classical results of Section II.

Our investigation into the physical origin of noise and dissipation in the
dynamics of the two point functions shows that in the final analysis this is
an effective dynamics, obtained from averaging out the higher correlations.
This point is made most explicit in the approach whereby the 2PI EA for the
correlations is obtained through truncation of the master effective action,
this being the formal functional whose variations generate the full
Schwinger - Dyson hierarchy. In Section V, we briefly discuss the definition
and construction of the master effective action, the relationship of
truncation to common approximation schemes, and present explicitly the
calculation leading to the dynamics of the two point functions at three
loops accuracy \cite{cddn,CH88}.

In the last section we give a brief discussion of the meaning of our results
and possible implications on renormalization group theory.

\section{Stochastic Boltzmann Equation from FDT}

As a primer, we wish to introduce the fluctuation-dissipation theorem (FDT)
or relation (FDR) in its simplest yet complete form, and apply it to derive
the stochastic Boltzmann equation so as to clarify its physical content.
There are many different versions \cite{BoonYip}: It could be taken to mean
the formulae relating dissipative coefficients to time integrals of
correlation functions (sometimes called the ``Landau - Lifshitz FDT) or the
relations between the susceptibility and the space integral of the
correlation function. In this paper, the fluctuation-dissipation theorem
addresses the relation between the dissipative coefficients of the effective
open system and the auto- correlation of random forces acting on the system,
as illustrated below.

\subsection{Fluctuation - dissipation theorem (FDT)}

The simplest setting \cite{Landau} for the FDT is a homogeneous system
described by variables $x^i$. The thermodynamics is encoded in the form of
the entropy $S\left( x^i\right) $. The thermodynamic fluxes are the
derivatives $\dot x^i$, and the thermodynamic forces are the components of
the gradient of the entropy

\begin{equation}
F_i=-\frac{\partial S}{\partial x^i}  \label{tf}
\end{equation}
The dynamics is given by

\begin{equation}
\dot x^i=-\gamma ^{ij}F_j+j^i  \label{langevin}
\end{equation}
The first term describes the mean regression of the system towards a local
entropy maximum, $\gamma^{ij}$ being the dissipative coefficient or
function, and the second term describes the random microscopic fluctuations
induced by its interaction with an environment. Near equilibrium, we also
have the phenomenological relations for linear response

\begin{equation}
F_i=c_{ij}x^j  \label{firstorder}
\end{equation}
where $c_{ij}$ is a nonsingular matrix.

In a classical theory, the equal time statistics of fluctuations is
determined by Einstein's law

\begin{equation}
\left\langle x^i\left( t\right) F_j\left( t\right) \right\rangle =\delta _j^i
\label{einstein}
\end{equation}
Take a derivative to find

\begin{equation}
0=c_{jk}\left\{ \left\langle \left( -\gamma ^{il}F_l+j^i\right)
x^k\right\rangle +\left\langle x^i\left( -\gamma ^{kl}F_l+j^k\right)
\right\rangle \right\}  \label{deriv}
\end{equation}
If the noise is gaussian,

\[
\left\langle x^i\left( t\right) j^k\left( t\right) \right\rangle =\int
dt^{\prime }\;\left. \frac{\delta x^i\left( t\right) }{\delta j^l\left(
t^{\prime }\right) }\right| \left\langle j^l\left( t^{\prime }\right)
j^k\left( t\right) \right\rangle 
\]
and white

\begin{equation}
\left\langle j^l\left( t^{\prime }\right) j^k\left( t\right) \right\rangle
=\nu ^{lk}\delta \left( t^{\prime }-t\right)  \label{corr}
\end{equation}
then

\begin{equation}
\left\langle x^i\left( t\right) j^k\left( t\right) \right\rangle =\frac 12%
\nu ^{ik}  \label{eqtime}
\end{equation}
From Eq. (\ref{deriv}) and (\ref{einstein}) we find the noise-noise
auto-correlation function $\nu ^{ik}$ is related to the symmetrized
dissipative function $\gamma ^{ik}$ by

\begin{equation}
\nu ^{ik}=\left[ \gamma ^{ik}+\gamma ^{ki}\right]  \label{fdt}
\end{equation}
which is the FDT in a simple classical formulation\footnote{%
To be concrete, this is the FDT {\it of the second kind} in the
classification of Ref. \cite{KuboII}. The FDT {\it of the first kind} is
further discussed in Ref. \cite{Jorge}. Also observe that we are only
concerned with small deviations from equilibrium; FDT's valid arbitrarily
far from equilibrium are discussed in Ref. \cite{Gallavotti}.}.

In the case of a one - dimensional system, the above argument can be
simplified even further because there is only one variable $x$, and $\gamma
, c,\nu $ are simply constants. In equilibrium, we have $\left\langle
x^2\right\rangle =c^{-1}.$ On the other hand, the late time solution of the
equations of motion reads

\[
x\left( t\right) =\int^tdu\;e^{-\gamma c\left( t-u\right) }j\left( u\right) 
\]
which implies $\left\langle x^2\right\rangle =\nu /2\gamma c$. Thus $\nu
=2\gamma $, in agreement with Eq. (\ref{fdt}).

\subsection{Boltzmann equation for a classical relativistic gas}

We shall apply the theory above to a dilute gas of relativistic classical
particles \cite{isr76}. The system is described by its one particle
distribution function $f\left( X,k\right) $, where $X$ is a position
variable, and $k$ is a momentum variable. Momentum is assumed to lie on a
mass shell $k^2+M^2=0$. (We use the MTW convention, with signature -+++ for
the background metric \cite{MTW}) and have positive energy $k^0>0$. In other
words, given a spatial element $d\Sigma ^\mu =n^\mu d\Sigma $ and a momentum
space element $d^4k$, the number of particles with momentum $k$ lying within
that phase space volume element is

\begin{equation}
dn=-4\pi f\left( X,k\right) \theta \left( k^0\right) \delta \left(
k^2+M^2\right) k^\mu n_\mu \;d\Sigma \frac{d^4k}{\left( 2\pi \right) ^4}
\label{distrif}
\end{equation}

The dynamics of the distribution function is given by the Boltzmann
equation, which we give in a notation adapted to our later needs, and for
the time being without the sought-after stochastic terms

\begin{equation}
k^\mu \frac \partial {\partial X^\mu }f\left( k\right) =I_{col}\left(
X,k\right)  \label{be}
\end{equation}

\begin{equation}
I_{col}=\frac{\lambda ^2}4\left( 2\pi \right) ^3\int \left[ \prod_{i=1}^3%
\frac{d^4p_i}{\left( 2\pi \right) ^4}\theta \left( p_i^0\right) \delta
\left( p_i^2+M^2\right) \right] \left[ \left( 2\pi \right) ^4\delta \left(
p_1+p_2-p_3-k\right) \right] {\bf I}  \label{icol}
\end{equation}

\begin{equation}
{\bf I}=\left\{ \left[ 1+f\left( p_3\right) \right] \left[ 1+f\left(
k\right) \right] f\left( p_1\right) f\left( p_2\right) -\left[ 1+f\left(
p_1\right) \right] \left[ 1+f\left( p_2\right) \right] f\left( p_3\right)
f\left( k\right) \right\}  \label{integ}
\end{equation}

The entropy flux is given by

\begin{equation}
S^\mu \left( X\right) =4\pi \int \frac{d^4p}{\left( 2\pi \right) ^4}\theta
\left( p^0\right) \delta \left( p^2+M^2\right) p^\mu \left\{ \left[
1+f\left( p\right) \right] \ln \left[ 1+f\left( p\right) \right] -f\left(
p\right) \ln f\left( p\right) \right\}  \label{entroflux}
\end{equation}
while the entropy itself $S$ is (minus) the integral of the flux over a
Cauchy surface. Now consider a small deviation from the equilibrium
distribution

\begin{equation}
f=f_{eq}+\delta f  \label{dev}
\end{equation}

\begin{equation}
f_{eq}=\frac 1{e^{\beta p^0}-1}  \label{feq}
\end{equation}
corresponding to the same particle and energy fluxes

\begin{equation}
\int \frac{d^4p}{\left( 2\pi \right) ^4}\theta \left( p^0\right) \delta
\left( p^2+M^2\right) p^\mu \delta f\left( p\right) =0  \label{partflux}
\end{equation}

\begin{equation}
\int \frac{d^4p}{\left( 2\pi \right) ^4}\theta \left( p^0\right) \delta
\left( p^2+M^2\right) p^\mu p^0\delta f\left( p\right) =0  \label{enerflux}
\end{equation}
Then the variation in entropy becomes

\begin{equation}
\delta S=-2\pi \int d^3X\int \frac{d^4p}{\left( 2\pi \right) ^4}\theta
\left( p^0\right) \delta \left( p^2+M^2\right) p^0\frac 1{\left[
1+f_{eq}\left( p\right) \right] f_{eq}\left( p\right) }\left( \delta
f\right) ^2  \label{deltaSa}
\end{equation}

In the classical theory, the distribution function is concentrated on the
positive frequency mass shell. Therefore, it is convenient to label momenta
just by its spatial components $\vec p,$ the temporal component being
necessarily $\omega _p=\sqrt{M^2+\vec p^2}>0.$ In the same way, it is
simplest to regard the distribution function as a function of the three
momentum $\vec p$ alone, according to the rule

\begin{equation}
f^{(3)}\left( X,\vec p\right) =f\left[ X,\left( \omega _p,\vec p\right)
\right]  \label{newf}
\end{equation}
where $f$ represents the distribution function as a function on four
dimensional momentum space, and $f^{\left( 3\right) }$ its restriction to
three dimensional mass shell. With this understood, we shall henceforth drop
the superscript, using the same symbol $f$ for both functions, since only
the distribution function on mass shell enters into our discussion. The
variation of the entropy now reads

\begin{equation}
\delta S=-\frac 12\int d^3X\int \frac{d^3p}{\left( 2\pi \right) ^3}\frac 1{%
\left[ 1+f_{eq}\left( p\right) \right] f_{eq}\left( p\right) }\left( \delta
f\right) ^2.  \label{deltaS}
\end{equation}
From Einstein's formula, we conclude that, in equilibrium, the distribution
function is subject to Gaussian fluctuations, with equal time mean square
value

\begin{equation}
\left\langle \delta f\left( t,\vec X,\vec p\right) \delta f\left( t,\vec Y,%
\vec q\right) \right\rangle =\left( 2\pi \right) ^3\delta \left( \vec X-\vec 
Y\right) \delta \left( \vec p-\vec q\right) \left[ 1+f_{eq}\left( p\right)
\right] f_{eq}\left( p\right)  \label{classcor}
\end{equation}

One of the goals of this paper is to rederive this result as the kinetic
theory limit of the general fluctuation formula given for the propagators in
the Introduction, Eq. (\ref{delta2}). For the time being, we only observe
that this fluctuation formula is quite independent of the processes which
sustain equilibrium; in particular, it holds equally for a free and an
interacting gas, since it contains no coupling constants.

In the interacting case, however, a stochastic source is necessary to
sustain these fluctuations. Following the discussion of the FDR above, we
compute these sources by writing the dissipative part of the equations of
motion in terms of the thermodynamic forces

\begin{equation}
F\left( X,\vec p\right) =\frac 1{\left[ 1+f_{eq}\left( p\right) \right]
f_{eq}\left( p\right) }\frac{\delta f\left( X,\vec p\right) }{\left( 2\pi
\right) ^3}  \label{thermoforce}
\end{equation}
To obtain an equation of motion for $f\left( X,\vec p\right) $ multiply both
sides of the Boltzmann equation Eq. (\ref{be}) by $\theta \left( k^0\right)
\delta \left( k^2+M^2\right) $ and integrate over $k^0$ to get

\begin{equation}
\frac{\partial f}{\partial t}+\frac{\vec k}{\omega _k}\vec \nabla f=\frac 1{%
\omega _k}I_{col}  \label{newbolt}
\end{equation}
Upon variation we get

\begin{equation}
\frac{\partial (\delta f)}{\partial t}+\frac{\vec k}{\omega _k}\vec \nabla
(\delta f)=\frac 1{\omega _k}\delta I_{col}  \label{linearb}
\end{equation}

When we write $\delta I_{col}$ in terms of the thermodynamic forces, we find
local terms proportional to $F\left( k\right) $ as well as nonlocal terms
where $F$ is evaluated elsewhere. We shall keep only the former, as it is
usually done in deriving the ``collision time approximation'' to the
Boltzmann equation \cite{Huang} (also related to the Krook - Bhatnager -
Gross kinetic equation), thus we write

\begin{equation}
\delta I_{col}\left( k\right) \sim -\omega _k\nu ^2(X,\vec k)F(X,\vec k)
\label{deltaicol}
\end{equation}
where

\begin{equation}
\nu ^2(X,\vec k)=\frac{\lambda ^2}{4\omega _k}\left( 2\pi \right) ^6\int
\left[ \prod_{i=1}^3\frac{d^4p_i}{\left( 2\pi \right) ^4}\theta \left(
p_i^0\right) \delta \left( p_i^2-M^2\right) \right] \left[ \left( 2\pi
\right) ^4\delta \left( p_1+p_2-p_3-k\right) \right] I_{+}  \label{sigmasq}
\end{equation}
$k^0=\omega _k$, and

\begin{equation}
I_{+}=\left[ 1+f_{eq}\left( p_1\right) \right] \left[ 1+f_{eq}\left(
p_2\right) \right] f_{eq}\left( p_3\right) f_{eq}\left( k\right)
\label{iplus}
\end{equation}

Among other things, the linearized form of the Boltzmann equation provides a
quick estimate of the relevant relaxation time. Let us assume the high
temperature limit, where $f\sim T/M$, and the integrals in Eq. (\ref{sigmasq}%
) are restricted to the range $p\leq M.$ Then simple dimensional analysis
yields the estimate $\tau \sim M/\lambda ^2T^2$ for the relaxation time
appropriate to long wavelength modes .

\subsection{Fluctuations in the Boltzmann equation}

Observance of the FDT demands that a stochastic source $j$ be present in the
Boltzmann equation Eq. (\ref{be}) (and its linearized form, Eq. (\ref
{linearb})) which should assume the Langevin form:

\begin{equation}
\frac{\partial f}{\partial t}+\frac{\vec k}{\omega _k}\vec \nabla f=\frac 1{%
\omega _k}I_{col}+j(X,\vec k)  \label{lanbe}
\end{equation}
Then

\begin{equation}
\left\langle j\left( X,\vec p\right) j\left( Y,\vec q\right) \right\rangle
=-\left\{ \frac 1{\omega _p}\frac{\delta I_{col}\left( X,\vec p\right) }{%
\delta F\left( Y,\vec q\right) }+\frac 1{\omega _q}\frac{\delta
I_{col}\left( Y,\vec q\right) }{\delta F\left( X,\vec p\right) }\right\}
\label{fdtbe}
\end{equation}

From Eqs. (\ref{deltaicol}), (\ref{sigmasq}) and (\ref{iplus}) we find the
noise auto-correlation 
\begin{equation}
\left\langle j(X,\vec k)j\left( Y,\vec p\right) \right\rangle =2\delta
^{\left( 4\right) }\left( X-Y\right) \delta \left( \vec k-\vec p\right) \nu
^2(X,\vec k)  \label{singucor}
\end{equation}
where $\nu ^2$ is given in Eq. (\ref{sigmasq}). Eqs. (\ref{singucor}) and (%
\ref{sigmasq}) are the solution to our problem, that is, they describe the
fluctuations in the Boltzmann equation, required by consistency with the
FDT. Observe that, unlike Eq. (\ref{classcor}), the mean square value of the
stochastic force vanishes for a free gas.

In this discussion, of course, we accepted the Boltzmann equation as given
without tracing its origin. We now want to see how the noises in Eq. (\ref
{singucor}) originate from a deeper level, that related to the higher
correlation functions, which we call the correlation noises.

\section{Kinetic field theory, from Dyson to Boltzmann}

Our goal in this section is to show how the Boltzmann equation arises as a
description of the dynamics of quasiparticles in the kinetic limit of field
theory. To this end, we shall adopt the view that the main element in the
description of a nonequilibrium quantum field is its Green functions, whose
dynamics is given by the Dyson equations. This connects with the results of
our earlier paper on dissipation in Boltzmann equations \cite{CH88}. The
task is to find the noise or fluctuation terms. The need to upgrade the
Boltzmann equation to a Langevin form will lead to a similar generalization
of Dyson's equations, whose physical origin will be the subject of the
remaining of the paper.

The discussion of propagators is simplest for a free field theory, and so,
following our choice of physical clarity over formal rigor in the
exposition, we shall first discuss nonequilibrium free fields. The general
case follows.

\subsection{Free fields and propagators}

Let us focus on the nonequilibrium dynamics of a real scalar quantum
(Heisenberg) field $\Phi \left( x\right) $, obeying the Klein - Gordon
equation

\begin{equation}
\left( \Box -m^2\right) \Phi \left( x\right) =0  \label{kg}
\end{equation}
and the canonical equal time commutation relations

\begin{equation}
\left[ \dot \Phi \left( \vec x,t\right) ,\Phi \left( \vec y,t\right) \right]
=-i\hbar \delta \left( \vec x-\vec y\right)  \label{etccr}
\end{equation}
(from here on, we take $\hbar =1$).

We shall assume throughout that the expectation value of the field vanishes.
Thus the simplest nontrivial description of the dynamics will be in terms of
the two - point or Green functions, namely the expectation values of various
products of two field operators. Of particular relevance is the Jordan
propagator

\begin{equation}
G\left( x,x^{\prime }\right) =\left\langle \left[ \Phi \left( x\right) ,\Phi
\left( x^{\prime }\right) \right] \right\rangle  \label{jordan}
\end{equation}
which for a free field is independent of the state of the field. From the
Jordan propagator we derive the causal propagators, advanced and retarded

\begin{equation}
G_{adv}\left( x,x^{\prime }\right) =-iG\left( x,x^{\prime }\right) \theta
\left( t^{\prime }-t\right) ,\quad G_{ret}\left( x,x^{\prime }\right)
=iG\left( x,x^{\prime }\right) \theta \left( t-t^{\prime }\right) \;
\label{adv}
\end{equation}
These propagators describe the evolution of small perturbations (they are
fundamental solutions to the Klein - Gordon equation) but contain no
information about the state. For that purpose we require other propagators,
such as the positive and negative frequency ones

\begin{equation}
G_{+}\left( x,x^{\prime }\right) =\left\langle \Phi \left( x\right) \Phi
\left( x^{\prime }\right) \right\rangle ,\quad G_{-}\left( x,x^{\prime
}\right) =\left\langle \Phi \left( x^{\prime }\right) \Phi \left( x\right)
\right\rangle  \label{g+}
\end{equation}
Observe that $G=G_{+}-G_{-}$. The symmetric combination gives the Hadamard
propagator

\begin{equation}
G_1=G_{+}+G_{-}=\left\langle \left\{ \Phi \left( x\right) ,\Phi \left(
x^{\prime }\right) \right\} \right\rangle  \label{hadamard}
\end{equation}

Note that while the Jordan, advanced and retarded propagators emphasize the
dynamics, and the negative, positive frequency and Hadamard propagators
emphasize the statistical aspects, two other propagators contain both kinds
of information. They are the Feynman and Dyson propagators

\begin{equation}
G_F\left( x,x^{\prime }\right) =\left\langle T\left[ \Phi \left( x\right)
\Phi \left( x^{\prime }\right) \right] \right\rangle =\frac 12\left[
G_1\left( x,x^{\prime }\right) +G\left( x,x^{\prime }\right) sign\left(
t-t^{\prime }\right) \right]  \label{feynman}
\end{equation}

\begin{equation}
G_D\left( x,x^{\prime }\right) =\left\langle \tilde T\left[ \Phi \left(
x\right) \Phi \left( x^{\prime }\right) \right] \right\rangle =\frac 12%
\left[ G_1\left( x,x^{\prime }\right) -G\left( x,x^{\prime }\right)
sign\left( t-t^{\prime }\right) \right]  \label{dyson}
\end{equation}
where $T$ stands for time-ordered product

\begin{equation}
T\left[ \Phi \left( x\right) \Phi \left( x^{\prime }\right) \right] =\Phi
\left( x\right) \Phi \left( x^{\prime }\right) \theta \left( t-t^{\prime
}\right) +\Phi \left( x^{\prime }\right) \Phi \left( x\right) \theta \left(
t^{\prime }-t\right)  \label{top}
\end{equation}
and $\tilde T$ for anti- temporal ordering

\begin{equation}
\tilde T\left[ \Phi \left( x\right) \Phi \left( x^{\prime }\right) \right]
=\Phi \left( x^{\prime }\right) \Phi \left( x\right) \theta \left(
t-t^{\prime }\right) +\Phi \left( x\right) \Phi \left( x^{\prime }\right)
\theta \left( t^{\prime }-t\right)  \label{atop}
\end{equation}

\subsection{Equilibrium structure of propagators}

In this subsection, we shall review several important properties of the
equilibrium propagators which follow from the KMS condition [Eq. (\ref{kms}]
below) \cite{KMS}, and general invariance properties.

In equilibrium, all propagators must be time-translation invariant, and may
be Fourier transformed

\begin{equation}
G\left( x,x^{\prime }\right) =\int \frac{d^4k}{\left( 2\pi \right) ^4}%
e^{ik\left( x-x^{\prime }\right) }G\left( k\right)  \label{fourier}
\end{equation}
In particular, because the Jordan propagator is antisymmetric, we must have $%
G\left( \omega ,\vec k\right) =-G\left( -\omega ,\vec k\right) .$ Also,
since $G\left( x,x^{\prime }\right) =G\left( x^{\prime },x\right)
^{*}=-G\left( x,x^{\prime }\right) ^{*}$, $G\left( k\right) =G\left(
k\right) ^{*}.$

The positive and negative frequency propagators are further related by the
KMS condition 
\begin{equation}
G_{+}\left[ \left( t,\vec x\right) ,\left( t^{\prime },\vec x^{\prime
}\right) \right] =G_{-}\left[ \left( t+i\beta ,\vec x\right) ,\left(
t^{\prime },\vec x^{\prime }\right) \right]  \label{kms}
\end{equation}
where $\beta $ is the inverse temperature. With $G_{+}-G_{-}=G$, we get

\begin{equation}
G_{+}\left( k\right) =\frac{G\left( k\right) }{1-e^{-\beta k^0}}=sign\left(
k^0\right) \left[ \theta \left( k^0\right) +\frac 1{e^{\beta \left|
k^0\right| }-1}\right] G\left( k\right)  \label{kms+}
\end{equation}

\begin{equation}
G_{-}\left( k\right) =\frac{G\left( k\right) }{e^{\beta k^0}-1}=sign\left(
k^0\right) \left[ \theta \left( -k^0\right) +\frac 1{e^{\beta \left|
k^0\right| }-1}\right] G\left( k\right)  \label{kms-}
\end{equation}
Adding these two equations, we find

\begin{equation}
G_1\left( k\right) =2sign\left( k^0\right) \left[ \frac 12+\frac 1{e^{\beta
\left| k^0\right| }-1}\right] G\left( k\right)  \label{kms1}
\end{equation}
We may consider this formula as the quantum generalization of the FDT, as we
shall see below. Let us stress that Eqs. (\ref{kms}) to (\ref{kms1}) hold
for interacting as well as free fields.

Of course, for the homogeneous solutions to the Klein - Gordon equation ($G$%
, $G_{+}$, $G_{-}$ and $G_1$) we must have

\begin{equation}
G(k)=\delta \left( k^2+m^2\right) g\left( k\right)   \label{massshella}
\end{equation}
which leads to

\begin{equation}
G_{ret}\left( x,x^{\prime }\right) =\int \frac{d^4k}{\left( 2\pi \right) ^4}%
\frac{e^{ik\left( x-x^{\prime }\right) }}{-\left( k^0-i\varepsilon \right)
^2+\omega _k^2}\left[ \frac{g\left( \omega _k,\vec k\right) }{2\pi }\right]
\label{grft}
\end{equation}
and to

\begin{equation}
G_F\left( x,x^{\prime }\right) =\int \frac{d^4k}{\left( 2\pi \right) ^4}%
e^{ik\left( x-x^{\prime }\right) }\left[ \frac{\left( -i\right) }{%
-k^0{}^2+\omega _k^2-i\varepsilon }+\frac{2\pi \delta \left( k^0{}^2-\omega
_k^2\right) }{e^{\beta \left| k^0\right| }-1}\right] \left[ \frac{g\left(
\omega _k,\vec k\right) }{2\pi }\right]  \label{gfft}
\end{equation}
with similar formulae for $G_{adv}$ and $G_D$, respectively. It is
remarkable that all propagators may be split into a vacuum and a thermal
contribution, with the thermal part being the same for all propagators
except $G$, $G_{ret}$ and $G_{adv}$, where it vanishes. Also, we have
expressed all propagators in terms of $g$; in the language of the Lehmann
decomposition, this is just the density of states \cite{bjorkendrell}.

We shall finish this subsection by expanding our remark on Eq. (\ref{kms1})
being the fluctuation dissipation theorem\cite{Heinzfdt}. Suppose we try to
explain the quantum and statistical fluctuations of the field by adding an
external source -$j\left( x\right) $ to the right hand side of the Klein -
Gordon equation (\ref{kg}). The resulting field would be

\[
\Phi \left( x\right) =\int d^4x^{\prime }\;G_{ret}\left( x,x^{\prime
}\right) j\left( x^{\prime }\right) 
\]
If the process is stationary

\begin{equation}
\left\langle j\left( x\right) j\left( x^{\prime }\right) \right\rangle =\int 
\frac{d^4k}{\left( 2\pi \right) ^4}e^{ik\left( x-x^{\prime }\right) }\nu
\left( k\right)  \label{jjcor}
\end{equation}
we get

\[
\nu \left( k\right) =\frac{G_1\left( k\right) }{2\left| G_{ret}\left(
k\right) \right| ^2} 
\]
From Eqs. (\ref{kms1}) and (\ref{grft})

\begin{equation}
\nu \left( k\right) =\left[ 1+\frac 2{e^{\beta \left| k^0\right| }-1}\right]
\left| ImG_{ret}^{-1}(k)\right|  \label{qFDT}
\end{equation}
which is a generalized form of the FDT, including both quantum and thermal
fluctuations.

So far, we have intentionally left everything expressed in terms of the
density of states $g\left( k\right) $. For a free field, we can compute this
explicitly

\begin{equation}
g\left( k\right) =2\pi sign\left( k^0\right)  \label{freegk}
\end{equation}
with which we can fill in the remaining results.

\subsection{Interacting fields and the Dyson equation}

Let us now consider a weakly interacting field, obeying the Heisenberg
equation

\begin{equation}
\left( \Box -m^2\right) \Phi \left( x\right) -\frac \lambda 6\Phi ^3\left(
x\right) =0  \label{lp4}
\end{equation}
and the same equal time canonical commutation relations, Eq. (\ref{etccr}).
As before, we shall assume that the expectation value of the field vanishes
identically, and seek to describe the dynamics in terms of the propagators
introduced earlier.

In the usual approach to field theory, where one focuses on computing the
S-matrix elements, rather than the causal evolution of fields, the leading
role is played by the Feynman propagator, which is directly related to the
S-matrix through the LSZ reduction formulae, and has a simple perturbative
expansion \cite{bjorkendrell,ItzykZuber,ramond}. We may obtain a dynamical
equation for the Feynman propagator by noting that, from Eq. (\ref{top})

\[
\left( -\Box +m^2\right) T\left[ \Phi \left( x\right) \Phi \left( x^{\prime
}\right) \right] =T\left[ \left( -\Box +m^2\right) \Phi \left( x\right) \Phi
\left( x^{\prime }\right) \right] -i\delta \left( t-t^{\prime }\right) 
\]
Therefore

\begin{equation}
\left( -\Box +m^2\right) G_F\left( x,x^{\prime }\right) =-i\delta \left(
t-t^{\prime }\right) -\frac \lambda 6\left\langle T\left[ \Phi ^3\left(
x\right) \Phi \left( x^{\prime }\right) \right] \right\rangle
\label{dysoninout}
\end{equation}
(cfr. Eq. (\ref{gfft})). This is the Dyson equation for the propagator,
relating the evolution of the Feynman propagator to higher order (in this
case, four point) correlation functions.

As different from an IN-OUT matrix element of the S-matrix, in this case we
have an IN - IN expectation value taken with respect to a nontrivial state
defined at some initial time. Thus the perturbative expansion of the self
energy term cannot be expressed in terms of the IN - IN Feynman propagator
alone. We should rather have

\[
\left\langle T\left[ \Phi ^3\left( x\right) \Phi \left( x^{\prime }\right)
\right] \right\rangle \sim 3G_F\left( x,x\right) G_F\left( x,x^{\prime
}\right) -i\lambda \int d^4y\left\{ G_F^3\left( x,y\right) G_F\left(
y,x^{\prime }\right) -G_{-}^3\left( x,y\right) G_{+}\left( y,x^{\prime
}\right) \right\} 
\]
(since we use full propagators in the internal lines, two-particle reducible
(2PR) graphs must not be included). Thus to obtain a self-consistent
dynamics, we must enlarge the set to include other propagators as well. Of
course, we are assuming that the initial state is such that Wick's theorem
holds (for example, that it is Gaussian) - this issue is discussed in detail
in \cite{CH88}. We are also leaving aside issues of renormalization \cite
{CH87}.

We overcome this difficulty by adopting as fundamental object the closed-
time-path ordered propagator $G_P\left( x^a,y^b\right) $. This object is
equivalent to four ordinary propagators: if we write $G_P\left(
x^a,y^b\right) =G^{ab}\left( x,y\right) $, then $G^{11}\left( x,y\right)
=G_F\left( x,y\right) $, $G^{12}\left( x,y\right) =G_{-}\left( x,y\right) $, 
$G^{21}\left( x,y\right) =G_{+}\left( x,y\right) $ and $G^{22}\left(
x,y\right) =G_D\left( x,y\right) .$

We can obtain closed dynamical equations for these four propagators.
Actually there is a slight redundancy, but this set has the advantage of
being very simple to handle (see Chou et al in \cite{ctp} for details). The
equations read

\begin{equation}
\left[ -\Box +m^2+\frac \lambda 2G_F(x,x)\right] G^{ab}(x,x^{\prime })-\frac{%
i\lambda ^2}6c_{cd}\int d^4y\;\Sigma ^{ac}(x,y)G^{db}\left( y,x^{\prime
}\right) =-ic^{ab}\delta \left( x-x^{\prime }\right)  \label{ctpeq}
\end{equation}

\begin{equation}
\Sigma ^{ac}(x,y)=\left[ G^{ac}(x,y)\right] ^3  \label{sigmaac}
\end{equation}

The matrix $c (c_{11}=c^{11}=1$, $c_{22}=c^{22}=-1,$ all others zero) keeps
track of the sign inversions associated with the reverse temporal ordering
of the second branch. This form of the Dyson equation is relevant to our
discussion.

\subsection{The kinetic theory limit}

In equilibrium the propagators are time-translation invariant. Out of
equilibrium this is no longer true. In the kinetic theory regime, however,
the propagators depend mostly on the difference variable $u=x-x^{\prime }$,
with the corresponding Fourier transform depending weakly on the center of
mass variable $X=\left( 1/2\right) (x+x^{\prime })$. As such, the
propagators take the form

\begin{equation}
G^{ab}(x,x^{\prime })=\int \frac{d^4k}{\left( 2\pi \right) ^4}e^{ik\left(
x-x^{\prime }\right) }G^{ab}\left( X,k\right)  \label{kinlim}
\end{equation}
The $\Sigma $ kernel has a similar expression

\[
\Sigma ^{ab}(x,x^{\prime })=\int \frac{d^4k}{\left( 2\pi \right) ^4}%
e^{ik\left( x-x^{\prime }\right) }\Sigma ^{ab}\left( X,k\right) 
\]

\begin{equation}
\Sigma ^{ab}\left( X,k\right) =\int \prod_{i=1}^3\left\{ \frac{d^4p_i}{%
\left( 2\pi \right) ^4}G^{ab}(X,p_i)\right\} \left[ \left( 2\pi \right)
^4\delta \left( \sum p_i-k\right) \right]  \label{skernel}
\end{equation}
The weak dependence on $X$ allows for the approximations (details in \cite
{CH88})

\[
G^{ab}(x,x)=\int \frac{d^4k}{\left( 2\pi \right) ^4}G^{ab}\left( x,k\right)
\sim \int \frac{d^4k}{\left( 2\pi \right) ^4}G^{ab}\left( X,k\right) 
\]

\[
\int d^4y\;\Sigma ^{ac}(x,y)G^{db}\left( y,x^{\prime }\right) \sim \int 
\frac{d^4k}{\left( 2\pi \right) ^4}e^{ik\left( x-x^{\prime }\right) }\Sigma
^{ac}\left( X,k\right) G^{db}\left( X,k\right) 
\]
and the equations of motion become

\begin{equation}
\left[ k^2-ik^\mu \frac \partial {\partial X^\mu }-\frac 14\Box _X+M^2\left(
X\right) \right] G^{ab}(X,k)-\frac{i\lambda ^2}6c_{cd}\;\Sigma
^{ac}(X,k)G^{db}\left( X,k\right) =-ic^{ab}  \label{klctpeq}
\end{equation}

\begin{equation}
M^2\left( X\right) =m^2+\frac \lambda 2\int \frac{d^4k}{\left( 2\pi \right)
^4}G^{ab}\left( X,k\right)  \label{physmass}
\end{equation}

Alternatively, we may think of the propagators as functions of $x^{\prime }$%
, leading to an equation of the form (cfr. Eq. (\ref{ctpeq}))

\begin{equation}
\left[ -\Box ^{\prime }+m^2+\frac \lambda 2G_F(x^{\prime },x^{\prime
})\right] G^{ab}(x,x^{\prime })-\frac{i\lambda ^2}6c_{cd}\int
d^4y\;G^{ac}(x,y)\Sigma ^{db}\left( y,x^{\prime }\right) =-ic^{ab}\delta
\left( x-x^{\prime }\right)  \label{ctpeqb}
\end{equation}
In the kinetic limit, this yields

\begin{equation}
\left[ k^2+ik^\mu \frac \partial {\partial X^\mu }-\frac 14\Box _X+M^2\left(
X\right) \right] G^{ab}(X,k)-\frac{i\lambda ^2}6c_{cd}\;G^{ac}(X,k)\Sigma
^{db}\left( X,k\right) =-ic^{ab}  \label{klctpeqb}
\end{equation}
Taking the average and the difference of Eqs. (\ref{klctpeq}) and (\ref
{klctpeqb}) we get

\begin{equation}
\left[ k^2-\frac 14\Box _X+M^2\left( X\right) \right] G^{ab}(X,k)-\frac{%
i\lambda ^2}{12}c_{cd}\;\left\{ \Sigma ^{ac}(X,k)G^{db}\left( X,k\right)
+G^{ac}(X,k)\Sigma ^{db}\left( X,k\right) \right\} =-ic^{ab}
\label{massshelcond}
\end{equation}

\begin{equation}
k^\mu \frac \partial {\partial X^\mu }G^{ab}(X,k)+\frac{\lambda ^2}{12}%
c_{cd}\;\left\{ \Sigma ^{ac}(X,k)G^{db}\left( X,k\right) -G^{ac}(X,k)\Sigma
^{db}\left( X,k\right) \right\} =0  \label{kineq}
\end{equation}
We recognize the first equation as a mass shell condition on the
nonequilibrium propagator. The second equation is the kinetic equation
proper, describing relaxation towards equilibrium.

To investigate further this equation, we observe that since both terms are
already of second order in $\lambda $ (see \cite{CH88}), it is enough to
solve the mass shell condition to zeroth order. That is, we assume that the
renormalized mass $M^2$ is actually position independent, and write

\begin{equation}
G^{ab}(X,k)=G_0^{ab}(M^2,k)+G_{stat}^{ab}(X,k)  \label{pd}
\end{equation}
where the $G_0^{ab}(M^2,k)$ are vacuum propagators for a free field with
mass $M^2$, and $G_{stat}$ is the non vacuum part

\begin{equation}
G_{stat}^{ab}(X,k)=2\pi \delta \left( k^2+M^2\right) f\left( X,k\right)
\label{f1pdf}
\end{equation}
which we assume is the same for all propagators involved, as in the free
field case. $f\left( X,k\right) $ has the physical interpretation of a one
particle distribution function for quasi particles built out of the field
excitations. Substituting Eqs. (\ref{pd}) and (\ref{f1pdf}) into (\ref{kineq}%
), and assuming, for example, that $k^0>0$ ($f$ must be even in $k$, because
of the symmetries of the propagators) immediately shows that the dynamics of 
$f$ is given by the Boltzmann equation (\ref{be}), (\ref{icol}) and (\ref
{integ}).

We shall not discuss further the region of validity of the hypothesis
underlying the kinetic limit, except to observe that this issue is far from
trivial. On general grounds, one expects that propagators will depend
strongly on the difference variable on scales $\tau _C\sim M^{-1}$. For
smooth initial conditions, the scale for dependence on the average variable
is set by the relaxation time $\tau \sim M/\lambda ^2T^2$. A nontrivial
kinetic limit exists if $\tau \gg \tau _C.$ Already this simple estimate
shows that one would expect trouble in theories with strictly massless
particles, such as gauge or Goldstone bosons \cite{mrow}. If particle masses
are not specially protected, then at large temperature the physical mass $%
M\sim \sqrt{\lambda }T,$ and $\tau _C/\tau \sim \lambda $ will in general be
suitably small.

\subsection{Stochastic Dyson equations}

Our derivation of kinetic theory from the Dyson equations leads to an
incomplete Boltzmann equation: in addition to the usual collision integral
an explicitly stochastic term ought to be in place. Moreover, fluctuation -
dissipation considerations demand that this stochastic term has
auto-correlation Eqs. (\ref{singucor}) and (\ref{sigmasq}). Since no
manipulation of the deterministic Dyson equations will yield a stochastic
term like this, we posit that when quantum field theory is viewed in the
statistical mechanical context, the Dyson equations themselves are
incomplete. Suppose we add a stochastic driving term $F^{ab}$ to them (we
shall justify this later):

\begin{equation}
\left[ -\Box +m^2+\frac \lambda 2G_F(x,x)\right] G^{ab}(x,x^{\prime })-\frac{%
i\lambda ^2}6c_{cd}\int d^4y\;\Sigma ^{ac}(x,y)G^{db}\left( y,x^{\prime
}\right) =-ic^{ab}\delta \left( x-x^{\prime }\right) -iF^{ab}(x,x^{\prime })
\label{stctpeq}
\end{equation}

\begin{equation}
\left[ -\Box ^{\prime }+m^2+\frac \lambda 2G_F(x^{\prime },x^{\prime
})\right] G^{ab}(x,x^{\prime })-\frac{i\lambda ^2}6c_{cd}\int
d^4y\;G^{ac}(x,y)\Sigma ^{db}\left( y,x^{\prime }\right) =-ic^{ab}\delta
\left( x-x^{\prime }\right) -i\tilde F^{ab}(x,x^{\prime })  \label{stctpeqb}
\end{equation}
In the kinetic limit, the random forces become

\begin{equation}
F^{ab}(x,x^{\prime })=\int \frac{d^4k}{\left( 2\pi \right) ^4}e^{ik\left(
x-x^{\prime }\right) }F^{ab}\left( X,k\right)  \label{klrf}
\end{equation}
(and similarly for $\tilde F$). Leaving aside the random fluctuations of the
mass shell, we find the new kinetic equation

\begin{equation}
k^\mu \frac \partial {\partial X^\mu }G^{ab}(X,k)+\frac{\lambda ^2}{12}%
c_{cd}\;\left\{ \Sigma ^{ac}(X,k)G^{db}\left( X,k\right) -G^{ac}(X,k)\Sigma
^{db}\left( X,k\right) \right\} =H^{ab}\left( X,k\right)  \label{stkineq}
\end{equation}
where

\begin{equation}
H^{ab}\equiv \frac 12\left[ F-\tilde F\right] ^{ab}\left( X,k\right)
\label{boltzmannforce}
\end{equation}
Our problem now is to justify changing the Dyson equation to Eq. (\ref
{stctpeq}), and to expound the physical meaning of this new stochastic
equation. To do this we need to use functional methods, to which we now turn.

\section{Correlation Noise and Stochastic Boltzmann Equation}

Our goal in this section is to show how noise terms such as those introduced
above from phenomenological considerations may actually be systematically
identified from an appropriate effective action. In this section we shall
limit ourselves to finding a suitable recipe to identify the noise terms,
and to compare the results to the phenomenological discussion above. The
physical foundations of the recipe shall be discussed in the following
section.

\subsection{Fluctuations in the propagators}

We shall now adapt the foregoing discussion to the study of fluctuations in
the dynamics of the two point functions. The first step is to notice that
this dynamics can be obtained from the variation of the 2PI action
functional (we derive this formula in Section V, Eq. (\ref{redloops}))

\begin{eqnarray}
\Gamma \left[ G^{ab}\right] &=&\frac{-i}2\ln \left[ DetG\right] -\frac 12%
c_{ab}\int d^4x\;\left( -\Box +m^2\right) G^{ab}\left( x,x\right)
\label{gamma2pib} \\
&&\ -\frac \lambda 8c_{abcd}\int d^4x\;G^{ab}\left( x,x\right) G^{ab}\left(
x,x\right)  \nonumber \\
&&\ +\frac{i\lambda ^2}{48}c_{abcd}c_{efgh}\int d^4xd^4x^{\prime
}\;G^{ae}\left( x,x^{\prime }\right) G^{bf}\left( x,x^{\prime }\right)
G^{cg}\left( x,x^{\prime }\right) G^{dh}\left( x,x^{\prime }\right) 
\nonumber
\end{eqnarray}
The resulting equations of motion

\begin{equation}
\frac{-i}2G_{ab}^{-1}-\frac 12\left[ c_{ab}\left( -\Box +m^2\right) +\frac 
\lambda 2c_{abcd}G^{cd}\left( x,x\right) \right] \delta \left( x,x^{\prime
}\right) +\frac{i\lambda ^2}{12}c_{ac}c_{bd}\left[ G^{cd}\left( x,x^{\prime
}\right) \right] ^3=0  \label{dyson2pi}
\end{equation}
are seen to be equivalent to the Dyson equations Eq. (\ref{ctpeq}).

As we discussed in the Introduction, we shall incorporate quantum
fluctuations in the evolution of the Green function $G^{ab}$ by explicitly
adding a stochastic source $(-1/2)\kappa _{ab}$ to the right hand side of
Eq. (\ref{dyson2pi}). Let us write ${\bf G}^{ab}=G^{ab}+\Delta ^{ab}$, and
expand the 2PI CTP EA to second order (the first order term vanishes by
virtue of Eq. (\ref{dyson2pi})),

\begin{equation}
\delta \Gamma =\delta _2\Gamma  \label{gammavar}
\end{equation}

\begin{eqnarray}
\delta _2\Gamma \left[ \Delta ^{ab}\right]  &=&\frac i4G_{ab}^{-1}\Delta
^{bc}G_{cd}^{-1}\Delta ^{da}-\frac \lambda 8c_{abcd}\int d^4x\;\Delta
^{ab}\left( x,x\right) \Delta ^{ab}\left( x,x\right)   \label{secondvara} \\
&&\ +\frac{i\lambda ^2}8c_{abcd}c_{efgh}\int d^4xd^4x^{\prime
}\;G^{ae}\left( x,x^{\prime }\right) G^{bf}\left( x,x^{\prime }\right)
\Delta ^{cg}\left( x,x^{\prime }\right) \Delta ^{dh}\left( x,x^{\prime
}\right)   \nonumber
\end{eqnarray}
From now on, we shall assume that the background tadpole vanishes, and
identify the mass with its renormalized value. We now have, as discussed in
the Introduction

\begin{equation}
\left\langle \Delta ^{ab}\Delta ^{cd}\right\rangle =i\left[ \frac{\delta
^2\Gamma }{\delta G^{ab}\delta G^{cd}}\right] ^{-1}  \label{delta2b}
\end{equation}
To sustain these fluctuations, the noise auto-correlation must be 
\[
\left\langle \kappa _{ab}\kappa _{cd}\right\rangle =\left( 4i\right) \left[
\left. \frac{\delta ^2}{\delta \Delta ^{ab}\delta \Delta ^{cd}}\delta
_2\Gamma \right| _{\Delta =0}\right] ^{\dagger } 
\]
That is

\begin{equation}
\left\langle \kappa _{ab}\left( x,x^{\prime }\right) \kappa _{cd}\left(
y,y^{\prime }\right) \right\rangle =N_{abcd}\left( x,x^{\prime },y,y^{\prime
}\right) +N_{abcd}^{int}\left( x,x^{\prime },y,y^{\prime }\right)
\label{mistery}
\end{equation}

\[
N_{abcd}\left( x,x^{\prime },y,y^{\prime }\right) =G_{da}^{-1}\left(
y^{\prime },x\right) G_{bc}^{-1}\left( x^{\prime },y\right)
+G_{ac}^{-1}\left( x,y\right) G_{db}^{-1}\left( y^{\prime },x^{\prime
}\right) 
\]
\begin{eqnarray*}
N_{abcd}^{int}\left( x,x^{\prime },y,y^{\prime }\right)  &=&\lambda ^2\left[
G^{eg}G^{fh}\right] \left( x,x^{\prime }\right) c_{acef}c_{bdgh}\delta
\left( x-y\right) \delta \left( x^{\prime }-y^{\prime }\right)  \\
&&-i\lambda c_{abcd}\delta \left( x-x^{\prime }\right) \delta \left(
y-y^{\prime }\right) 
\end{eqnarray*}

\subsection{Free fields}

Let us begin by asking whether for free fields the quantum fluctuations Eq. (%
\ref{delta2b}) go into anything like the classical result Eq. (\ref{classcor}%
) in the kinetic theory limit. There is no obvious reason why it should be
so, since the physical basis for either formula is at first sight totally
different. As we saw in the Introduction, Eq. (\ref{delta2b}) simply
reproduces the full quantum fluctuations, computed in terms of the
propagators themselves on the assumption that Wick theorem holds (which is
an assumption on the allowed initial states of the field, see \cite{CH88})

\begin{equation}
\left\langle \Delta ^{ab}\Delta ^{cd}\right\rangle =i\left[ \frac{\delta
^2\Gamma }{\delta G^{ab}\delta G^{cd}}\right] ^{-1}=G^{ac}G^{db}+G^{da}G^{bc}
\label{freefluc2}
\end{equation}
while the classical auto-correlation Eq. (\ref{classcor}) has been found by
applying Einstein's formula to the phenomenological entropy eq. (\ref
{entroflux}). The only clear point of contact between both approaches is
that both assume Bose statistics.

Introducing the Wigner transform of the fluctuations

\begin{equation}
\Delta ^{ab}\left( x,x^{\prime }\right) =\int \frac{d^4k}{\left( 2\pi
\right) ^4}e^{ik\left( x-x^{\prime }\right) }\Delta ^{ab}\left( X,k\right)
;\qquad X=\frac 12\left( x+x^{\prime }\right).  \label{wignerdelta}
\end{equation}
we observe that in this case we are not entitled to assume that the
dependence of the Wigner transform on $X$ is weak. Eq. (\ref{wignerdelta})
has a formal inverse

\begin{equation}
\Delta ^{ab}\left( X,k\right) =\int du\;e^{iku}\Delta ^{ab}\left( X+\frac u2%
,X-\frac u2\right),  \label{formalinv}
\end{equation}
and from Eq. (\ref{freefluc2}) we get

\[
\left\langle \Delta ^{ab}\left( X,p\right) \Delta ^{cd}\left( Y,q\right)
\right\rangle =\int dudv\;e^{i\left( pu+qv\right) }K\left[ X,Y,u,v\right] , 
\]
where

\[
K\left[ X,Y,u,v\right] =G^{ac}\left( X+\frac u2,Y+\frac v2\right)
G^{bd}\left( X-\frac u2,Y-\frac v2\right) +G^{ad}\left( X+\frac u2,Y-\frac v2%
\right) G^{bc}\left( X-\frac u2,Y+\frac v2\right) . 
\]

The propagators in the right hand side are equilibrium ones, and so we can
use the representation Eq. (\ref{fourier})

\[
K\left[ X,Y,u,v\right] =\int \frac{d^4r}{\left( 2\pi \right) ^4}\frac{d^4s}{%
\left( 2\pi \right) ^4}\int dudv\;e^{i\left( pu+qv\right) }K\left[
X,Y,r,s\right] 
\]

\begin{eqnarray*}
K\left[ X,Y,r,s\right]  &=&\ \ e^{ir\left( X-Y+\frac 12\left( u-v\right)
\right) }e^{is\left( X-Y-\frac 12\left( u-v\right) \right) }G^{ac}\left(
r\right) G^{bd}\left( s\right)  \\
&&+e^{ir\left( X-Y+\frac 12\left( u+v\right) \right) }e^{is\left( X-Y-\frac 1%
2\left( u+v\right) \right) }G^{ad}\left( r\right) G^{bc}\left( s\right) 
\end{eqnarray*}
Now integrate over $u$, $v$ and $s$

\begin{eqnarray}
\left\langle \Delta ^{ab}\left( X,p\right) \Delta ^{cd}\left( Y,q\right)
\right\rangle &=&16\int d^4r\;e^{i2\left( r+p\right) \left( X-Y\right) }
\label{quantumcor} \\
&&\ \ \ \ \left[ \delta \left( p+q\right) G^{ac}\left( r\right) G^{bd}\left(
r+2p\right) +\delta \left( p-q\right) G^{ad}\left( r\right) G^{bc}\left(
r+2p\right) \right]  \nonumber
\end{eqnarray}
We have 16 different quantum auto-correlations to compare against a single
classical result, so we can only expect real agreement in the large
occupation number limit, where all propagators converge to the same
expression. With this proviso in mind, we can choose any combination of
indices to continue the calculation. The most straightforward choice (to a
certain extent suggested by the structure of the closed time - path, see 
\cite{dch,cddn}) is $a=b=1$, $c=d=2$; we are thus seeking the correlations
among the fluctuations in the Feynman and Dyson propagators

\begin{eqnarray*}
\left\langle \Delta ^{11}\left( X,p\right) \Delta ^{22}\left( Y,q\right)
\right\rangle &=&16\left( 2\pi \right) ^2\left[ \delta \left( p+q\right)
+\delta \left( p-q\right) \right] \int d^4r\;e^{i2\left( r+p\right) \left(
X-Y\right) }\delta \left[ r^2+M^2\right] \delta \left[ \left( r+2p\right)
^2+M^2\right]  \label{quantumcorb} \\
&&\left[ \theta \left( -r^0\right) +f_{eq}\left( r\right) \right] \left[
\theta \left( -r^0-2p^0\right) +f_{eq}\left( r+2p\right) \right] \ \ \ \ \ \
\ 
\end{eqnarray*}
The arguments of the delta functions can be simplified

\begin{eqnarray}
\left\langle \Delta ^{11}\left( X,p\right) \Delta ^{22}\left( Y,q\right)
\right\rangle  &=&4\left( 2\pi \right) ^2\left[ \delta \left( p+q\right)
+\delta \left( p-q\right) \right] \int d^4r\;e^{i2\left( r+p\right) \left(
X-Y\right) }\delta \left[ r^2+M^2\right]   \label{quantumcorb} \\
&&\delta \left[ rp+p^2\right] \left[ \theta \left( -r^0\right) +f_{eq}\left(
r\right) \right] \left[ \theta \left( -r^0-2p^0\right) +f_{eq}\left(
r+2p\right) \right]   \nonumber
\end{eqnarray}

A difference from the classical case already stands out here: in the quantum
case, a fluctuation in the number of particles with momentum $p$ correlates
not only with itself, but also with the corresponding fluctuation in the
number of antiparticles with momentum $-p$. This is unavoidable, given the
symmetries of the propagators in this theory.

Let us stress that we are trying to push the quasi - particle (kinetic)
description of quantum field dynamics beyond the calculation of mean values
(of such quantities as particle number or energy density), to account for
their fluctuations. The calculation of the fluctuations of the distribution
function for on-shell particles gives a crucial consistency check on such an
attempt. Indeed, we know that each on-shell mode of the free field
contributes an amount [cfr. Eqs. (\ref{kms1}), (\ref{massshell}) and (\ref
{freegk})] $\rho _k\sim \omega _k\left( 1/2+f_{eq}\right) $to the mean
energy density, where $f_{eq}$ is the equilibrium distribution function Eq. (%
\ref{feq}). The fluctuations of this quantity at equilibrium will be given
by $\left\langle \delta \rho _k^2\right\rangle =T^2\left( \partial \rho
_k/\partial T\right) \sim \omega _k^2f_{eq}.\left( 1+f_{eq}\right) $. So, 
{\it if} these fluctuations are still described by a distribution function
consistent with ordinary statistical mechanics, then this distribution
function {\it must} fluctuate like in Eq. (\ref{classcor}). (This may in the
face of it be a rather big {\it \ if}).

For large $M^2$, the condition that $p$ is nearly on - shell means that the
spatial components are much smaller than the time component, and we may
approximate

\[
\delta \left[ r^2+M^2\right] \delta \left[ rp+p^2\right] \sim \delta \left[
p^2+M^2\right] \frac 1{\left| p^0\right| }\delta \left( r^0+p^0\right) 
\]
thus obtaining

\[
\left. \left\langle \Delta ^{11}\left( X,p\right) \Delta ^{22}\left(
Y,q\right) \right\rangle \right| _{{\rm on-shell}}=\frac 1{2\omega _p}\left(
2\pi \right) ^5\left[ \delta \left( p+q\right) +\delta \left( p-q\right)
\right] f_{eq}\left( p\right) \left( 1+f_{eq}\left( p\right) \right) \delta
\left[ p^2+M^2\right] \delta \left( \vec X-\vec Y\right) 
\]

To finish the comparison, assume, e. g., that $p^0\geq 0$, then

\begin{eqnarray}
\delta \left( p-q\right) \delta \left[ p^2+M^2\right] &=&\delta \left(
q^0-\omega _q\right) \delta \left( \vec q-\vec p\right) \delta \left[
p^2+M^2\right]  \label{massshell} \\
&=&2\omega _q\theta \left( q^0\right) \delta \left( \vec q-\vec p\right)
\delta \left[ p^2+M^2\right] \delta \left[ q^2+M^2\right]  \nonumber
\end{eqnarray}
This results suggests writing

\begin{equation}
\Delta ^{11}\left( X,p\right) =2\pi \delta f\left( X,\vec p\right) \delta
\left[ p^2+M^2\right] +{\rm off-shell\;terms}.  \label{onshellfluc}
\end{equation}
Taking $p^0$ and $q^0$ to be positive, this yields

\begin{equation}
\left\langle \delta f\left( t,\vec X,\vec p\right) \delta f\left( t,\vec Y,%
\vec q\right) \right\rangle =\left( 2\pi \right) ^3\delta \left( \vec q-\vec 
p\right) \delta \left( \vec X-\vec Y\right) f_{eq}\left( p\right) \left(
1+f_{eq}\left( p\right) \right)  \label{quantumcorc}
\end{equation}
which is identical to Eq. (\ref{classcor}). This is one of the most
important results of this paper, as it gives a whole new meaning to the
phenomenological entropy Eq. (\ref{entroflux})

We have thus completed our proof, and obtained new independent confirmation
of the validity of our scheme for introducing fluctuations in the dynamics
of correlations.

\subsection{Interacting fields and the Boltzmann - Langevin equation}

The results of the previous section already imply that the full stochastic
Dyson equation will go over to the Boltzmann - Langevin equation in the
kinetic limit. Indeed, the structure of the fluctuations does not change
drastically when interactions are switch on, and since they become identical
in the classical limit, the noise in the Dyson equation necessary to sustain
the fluctuations at the quantum level must go over to the noise in the
Boltzmann equation, which plays the same role in the classical theory.
Nevertheless, it is worth identifying exactly which part of the quantum
source auto - correlation goes into the classical one in the correspondence
limit.

Concretely, our aim is to begin with the stochastic Schwinger - Dyson
equation

\begin{equation}
\frac{-i}2{\bf G}_{ab}^{-1}-\frac 12\left[ c_{ab}\left( -\Box +m^2\right) +%
\frac \lambda 2c_{abcd}{\bf G}^{cd}\left( x,x\right) \right] \delta \left(
x,x^{\prime }\right) +\frac{i\lambda ^2}{12}c_{ac}c_{bd}\left[ {\bf G}%
^{cd}\left( x,x^{\prime }\right) \right] ^3=\frac{-1}2\kappa _{ab}
\label{stdyson2pi}
\end{equation}
where the noise auto - correlation is given by Eq. (\ref{mistery}). We then
identify the forces appearing in Eqs. (\ref{stctpeq}) and (\ref{stctpeqb})

\begin{equation}
F^{ab}(x,x^{\prime })=i\int d^4y\;c^{ac}\kappa _{cd}\left( x,y\right)
G^{db}\left( y,x^{\prime }\right)  \label{ff}
\end{equation}

\[
\tilde F^{ab}(x,x^{\prime })=i\int d^4y\;G^{ac}\left( x,y\right) \kappa
_{cd}\left( y,x^{\prime }\right) c^{db} 
\]
In condensed notation,

\begin{equation}
H^{ab}\equiv \frac 12\left[ F-\tilde F\right] ^{ab}=\frac i2\left\{
c^{ac}\kappa _{cd}G^{db}-G^{ac}\kappa _{cd}c^{db}\right\}  \label{hh}
\end{equation}
whose Wigner transform plays the role of random force in the kinetic
equation (\ref{stkineq}). Restricting ourselves to on-shell fluctuations, we
can compute the auto-correlation of this force and compare the result to the
classical expectation Eq. (\ref{singucor}).

Let us observe from the outset that the classical result involves the
expression $\nu ^2$ (Eq. (\ref{sigmasq})), which, through $I_{+}$ (Eq.(\ref
{iplus})), is related to the Fourier transform of the cube of a propagator.
In Eq. (\ref{mistery}), the first term $N$ contains the inverse propagators,
which in turn is related to the cube of propagators through the Dyson
equations (\ref{dyson2pi}). The other term $N^{int}$ contains no such thing.
Thus it is clear that our only chance lies in the first term, the other one
contributing to sustaining the nonclassical correlations already present in
the free field case. Correspondingly, we shall ignore $N^{int}$ in what
follows.

We thus approximate

\begin{equation}
\left\langle \kappa _{ab}\kappa _{cd}\right\rangle
=G_{da}^{-1}G_{bc}^{-1}+G_{ac}^{-1}G_{db}^{-1}  \label{mistery2}
\end{equation}
leading to

\begin{eqnarray*}
\left\langle H^{ab}H^{cd}\right\rangle &=&\frac{-1}4\left[
c^{ae}G^{fb}-G^{ae}c^{fb}\right] \left[ c^{cg}G^{hd}-G^{cg}c^{hd}\right]
\left[ G_{he}^{-1}G_{fg}^{-1}+G_{eg}^{-1}G_{hf}^{-1}\right] \\
&=&\frac 14\left[ -c^{ae}G^{fb}+G^{ae}c^{fb}\right] \left[ \delta
_e^dG_f^{-1c}+\delta _f^dG_e^{-1c}-\delta _f^cG_e^{-1d}-\delta
_e^cG_f^{-1d}\right] \\
&=&\frac 14\left[
G^{ad}G^{-1bc}+G^{-1ad}G^{bc}-G^{bd}G^{-1ac}-G^{-1bd}G^{ac}+2\left(
c^{ac}c^{bd}-c^{ad}c^{bc}\right) \right]
\end{eqnarray*}
For the same reasons as above, we shall disregard the propagator -
independent terms.

Next, we write (recalling Eq. (\ref{sigmaac}))

\begin{equation}
H^{ab}(x,x^{\prime })=\int \frac{d^4k}{\left( 2\pi \right) ^4}e^{ik\left(
x-x^{\prime }\right) }H^{ab}\left( X,k\right)  \label{klrh}
\end{equation}

\[
H^{ab}\left( X,k\right) =\int d^4u\;e^{-iku}H^{ab}\left( X+\frac u2,X-\frac u%
2\right) 
\]
to get

\begin{equation}
\left\langle H^{ab}\left( X,p\right) H^{cd}\left( Y,q\right) \right\rangle =%
\frac 14\int d^4ud^4v\;e^{-i\left( pu+qv\right) }\left[ J-K\right],
\end{equation}
where, using the translation invariance of the equilibrium propagators

\begin{eqnarray*}
J &=&\int \frac{d^4r}{\left( 2\pi \right) ^4}\frac{d^4s}{\left( 2\pi \right)
^4}\;\exp \left[ ir\left( X-Y+\frac{u+v}2\right) \right] \exp \left[
is\left( X-Y-\frac{u+v}2\right) \right]  \\
&&\left\{ G^{ad}\left( r\right) G^{-1bc}\left( s\right) +G^{-1ad}\left(
r\right) G^{bc}\left( s\right) \right\} 
\end{eqnarray*}

\begin{eqnarray*}
K &=&\int \frac{d^4r}{\left( 2\pi \right) ^4}\frac{d^4s}{\left( 2\pi \right)
^4}\;\exp \left[ ir\left( X-Y-\frac{u-v}2\right) \right] \exp \left[
is\left( X-Y+\frac{u-v}2\right) \right]  \\
&&\left\{ G^{bd}\left( r\right) G^{-1ac}\left( s\right) +G^{-1bd}\left(
r\right) G^{ac}\left( s\right) \right\} .
\end{eqnarray*}

Upon integration over $u$ and $v$, the $K$ term gives a contribution
proportional to $\delta ^{(4)}\left( p+q\right) $. This is unrelated to the
noise auto - correlation, being only a cross correlation between the
positive and negative frequency components of the source, and we shall not
analyze it further. We also restrict ourselves to the case where $a=b=1$, $%
c=d=2$. Using the reflection symmetry of the equilibrium propagators,
obtaining the inverse propagators from Eq. (\ref{dyson2pi}) and retaining
only the dominant term in the correspondence limit, we find

\begin{equation}
\left\langle H^{11}\left( X,p\right) H^{22}\left( Y,q\right) \right\rangle
\sim \frac{4\lambda ^2}3\delta \left( p-q\right) \int d^4s\;\cos \left[
2\left( s+p\right) \left( X-Y\right) \right] \Sigma ^{12}\left( s+2p\right)
G^{12}\left( s\right).  \label{hcor}
\end{equation}

An analysis of this expression shows that in the high temperature limit the
correlation length is of order $M^{-1}$. This is a microscopic scale, much
smaller than the macro scales of relevance to the kinetic limit (if this
limit exists). Therefore we are justified in writing

\begin{equation}
\left\langle H^{11}\left( X,p\right) H^{22}\left( Y,q\right) \right\rangle
\sim \gamma \delta \left( X-Y\right)  \label{macrocor}
\end{equation}
We compute $\gamma $ by simply integrating Eq. (\ref{hcor}) over $X$

\begin{equation}
\gamma =\frac{\lambda ^2}{12}\left( 2\pi \right) ^4\Sigma ^{12}\left(
p\right) G^{21}\left( p\right) \delta \left( p-q\right) .
\end{equation}

From Eqs. (\ref{skernel}) and (\ref{sigmasq}) we get

\begin{equation}
\gamma =\omega _p\left( 2\pi \right) ^2\delta \left( p-q\right) \delta
\left( p^2+M^2\right) \nu ^2
\end{equation}

Assuming $p^0$, $q^0\geq 0$, this is

\begin{equation}
\gamma =2\omega _p^2\left( 2\pi \right) ^2\delta \left( \vec p-\vec q\right)
\delta \left( q^2+M^2\right) \delta \left( p^2+M^2\right) \nu ^2
\end{equation}

So, writing

\begin{equation}
H\left( X,k\right) =2\pi \omega _k\delta \left( k^2+M^2\right) j\left( X,%
\vec k\right) +{\rm off-shell}
\end{equation}

we get the final result

\begin{equation}
\left\langle j\left( X,\vec p\right) j\left( Y,\vec q\right) \right\rangle
=2\delta \left( \vec p-\vec q\right) \delta \left( X-Y\right) \nu ^2,
\end{equation}

which agrees with the classical result, Eq. (\ref{singucor}).

We have shown that there is a piece of the full quantum noise which can be
identified with the classical source $j$. Clearly $j$ does not account for
the full quantum noise, the difference being due among other things to the
role of negative frequency in the quantum theory.

Finally, we note that Abe {\it et al.}\cite{abe}{\it \ }have given a
nonrelativistic derivation of the Boltzmann - Langevin equation, while ours
is fully relativistic, being also immune to the reservations expressed by
Greiner and Leupold \cite{greinleu}.

\section{Master Effective Action}

So far in the paper we have referred several times to the possibility of
conceiving the low order correlations of a quantum field as the field
variables of an open quantum system, interacting with the environment
provided by the higher correlations. The goal of this final section is to
present a formalism, the master effective action, built on this perspective.
In particular, in this formalism the usual Dyson equations are seen to
emerge from the averaging over higher correlations. As a simple consequence
and illustration, we derive Eq. (\ref{gamma2pib}).

\subsection{The low order effective actions}

The simplest application of functional methods in quantum field theory
concerns the dynamics of the expectation value of the field \cite{effaction}%
. The expectation value or mean field may be deduced from the generating
functional $W\left[ J\right] $

\begin{equation}
\exp \left\{ iW\left[ J\right] \right\} =\int D\Phi \;\exp \left\{ iS\left[
\Phi \right] +i\int d^4x\;J\left( x\right) \Phi \left( x\right) \right\} 
\label{genfuna}
\end{equation}

\begin{equation}
\phi \left( x\right) =\left. \frac{\delta W}{\delta J}\right| _{J=0}
\label{orpar}
\end{equation}
We obtain the dynamics from the effective action, which is the Legendre
transform of $W$

\begin{equation}
\Gamma \left[ \phi \right] =W\left[ J\right] -\int d^4x\;J\left( x\right)
\phi \left( x\right)  \label{efact}
\end{equation}
The physical equation of motion is

\begin{equation}
\frac{\delta \Gamma }{\delta \phi }=0  \label{eqmot}
\end{equation}

In a causal theory, we must adopt Schwinger's CTP formalism. The point $x$
may therefore lie on either branch of the closed time path, or equivalently
we may have two background fields $\phi ^a\left( x\right) =\phi \left(
x^a\right) $. The classical action is defined as

\begin{equation}
S\left[ \Phi ^a\right] =S\left[ \Phi ^1\right] -S\left[ \Phi ^2\right] ^{*}
\label{ctpclac}
\end{equation}
which automatically accounts for all sign reversals. We also have two sources

\[
\int d^4x\;J_a\left( x\right) \Phi ^a\left( x\right) =\int d^4x\;\left[
J^1\left( x\right) \Phi ^1\left( x\right) -J^2\left( x\right) \Phi ^2\left(
x\right) \right] 
\]
and obtain two equations of motion

\begin{equation}
\frac{\delta \Gamma }{\delta \phi ^a}=0  \label{ctpeqmot}
\end{equation}
However, these equations always admit a solution where $\phi ^1=\phi ^2=\phi 
$ is the physical mean field, and after this identification, they become a
real and causal equation of motion for $\phi $.

The functional methods we have used so far to derive the dynamics of the
mean field may be adapted to investigate more general operators. In order to
find the equations of motion for two-point functions, for example, we add a
nonlocal source $K_{ab}(x,x^{\prime })$\cite{pi2,CH88}

\begin{equation}
\exp \left\{ iW\left[ J_a,K_{ab}\right] \right\} =\int D\Phi ^a\;\exp
i\left\{ S\left[ \Phi ^a\right] +\int d^4x\;J_a\Phi ^a+\frac 12\int
d^4xd^4x^{\prime }\;K_{ab}\Phi ^a\Phi ^b\right\}  \label{tpigf}
\end{equation}
It follows that

\[
\frac{\delta W}{\delta K_{ab}\left( x,x^{\prime }\right) }=\frac 12\left[
\phi ^a\left( x\right) \phi ^b\left( x^{\prime }\right) +G^{ab}\left(
x,x^{\prime }\right) \right] 
\]
Therefore the Legendre transform, the so-called 2PI effective action,

\begin{equation}
\Gamma \left[ \phi ^a,G^{ab}\right] =W\left[ J_a,K_{ab}\right] -\int
d^4x\;J_a\phi ^a-\frac 12\int d^4xd^4x^{\prime }\;K_{ab}\left[ \phi ^a\phi
^b+G^{ab}\right]  \label{gamma2pi}
\end{equation}
generates the equations of motion

\begin{equation}
\frac{\delta \Gamma }{\delta \phi ^a}=-J_a-K_{ab}\phi ^b;\;\frac{\delta
\Gamma }{\delta G^{ab}}=-\frac 12K_{ab}  \label{funceqmot}
\end{equation}

The goal of this section is to show these two examples as just successive
truncations of a single object, the master effective action.

\subsection{Formal construction}

In this section, we shall proceed with the formal construction of the master
effective action, a functional of the whole string of Green functions of a
field theory whose variation generates the Dyson - Schwinger hierarchy.
Since we are using Schwinger - Keldish techniques, all fields are to be
defined on a closed time path. Also we adopt DeWitt's condensed notation\cite
{DeWitt}.

We consider then a scalar field theory whose action

\begin{equation}
S[\Phi ]={\frac{1}{2}}S_2\Phi^2 +S_{int}[\Phi ]
\end{equation}
decomposes into a free part and an interaction part

\begin{equation}
S_{int}[\Phi ]=\sum_{n=3}^{\infty}{\frac{1}{n!}}S_n\Phi^n
\end{equation}
Here and after, we use the shorthand

\begin{equation}
K_n\Phi^n\equiv\int~d^dx_1...d^dx_n~K_{na^1...a^n}(x_1,...x_n)
\Phi^{a^1}(x_1)...\Phi^{a^n}(x_n)
\end{equation}
where the kernel $K$ is assumed to be totally symmetric.

Let us define also the `source action'

\begin{equation}
J[\Phi ]=J_1\Phi +{\frac{1}{2}}J_2\Phi^2+J_{int}[\Phi ]
\end{equation}
where $J_{int}[\Phi ]$ contains the higher order sources

\begin{equation}
J_{int}[\Phi ]=\sum_{n=3}^{\infty}{\frac{1}{n!}}J_n\Phi^n
\end{equation}
and define the generating functional

\begin{equation}
Z[\{J_n\}]=e^{iW[\{J_n\}]}=\int~D\Phi~e^{iS_t[\Phi ,\{J_n\}]}
\end{equation}
where

\begin{equation}
S_t[\Phi ,\{J_n\}]=J_1\Phi +{\frac{1}{2}}(S_2+J_2)\Phi^2+S_{int}[\Phi ]
+J_{int}[\Phi ]
\end{equation}
We shall also call

\begin{equation}
S_{int}[\Phi ]+J_{int}[\Phi ]=S_I
\end{equation}

As it is well known, the Taylor expansion of $Z$ with respect to $J_1$
generates the expectation values of path - ordered products of fields

\begin{equation}
{\frac{\delta^n Z}{\delta J_{1a^1}(x_1)...\delta J_{1a^n}(x_n)}}= \langle
P\{\Phi^{a^1}(x_1)...\Phi^{a^n}(x_n)\}\rangle\equiv F_n^{a^1...a^n}
(x_1,...x_n)
\end{equation}
while the Taylor expansion of $W$ generates the `connected' Green functions
(`linked cluster theorem' \cite{Haag})

\begin{equation}
{\frac{\delta^n W}{\delta J_{1a^1}(x_1)...\delta J_{1a^n}(x_n)}}= \langle
P\{\Phi^{a^1}(x_1)...\Phi^{a^n}(x_n)\}\rangle_{{\rm connected}}\equiv
C_n^{a^1...a^n} (x_1,...x_n)
\end{equation}
Comparing these last two equations, we find the rule connecting the $F$'s
with the $C$'s. First, we must decompose the ordered index set $(i_1,...i_n)$
($i_k=(x_k,a^k)$) into all possible clusters $P_n$. A cluster is a partition
of $(i_1,...i_n)$ into $N_{P_n}$ ordered subsets $p=(j_1,...j_r)$. Then

\begin{equation}
F_n^{i_1...i_n}=\sum_{P_n}\prod_{p}C_r^{j_1...j_r}
\end{equation}
Now from the obvious identity

\begin{equation}
{\frac{\delta Z}{\delta J_{ni_1...i_n}}}\equiv \frac{1}{n!} {\frac{\delta^n Z%
}{\delta J_{i_1}...\delta J_{i_n}}}
\end{equation}
we obtain the chain of equations

\begin{equation}
{\frac{\delta W}{\delta J_{ni_1...i_n}}}\equiv {\frac{1}{n!}}
\sum_{P_n}\prod_{p}C_r^{j_1...j_r}
\end{equation}

We can invert these equations to express the sources as functionals of the
connected Green functions, and define the master effective action (MEA) as
the full Legendre transform of the connected generating functional

\begin{equation}
\Gamma_{\infty}[\{C_r\}]=W[\{J_n\}]-\sum_n{\frac{1}{n!}}J_n\sum_{P_n}%
\prod_pC_r
\end{equation}
The physical case corresponds to the absence of external sources, whereby

\begin{equation}
{\frac{\delta\Gamma_{\infty}[\{C_r\}]}{\delta C_s}}=0
\end{equation}
This hierarchy of equations is equivalent to the Dyson- Schwinger series.

\subsection{The background field method}

The master effective action just introduced becomes more manageable if one
applies the background field method (BFM) \cite{effaction} approach. We
first distinguish the mean field and the two point functions

\begin{equation}
C_1^i\equiv\phi^i
\end{equation}

\begin{equation}
C_2^{ij}\equiv G^{ij}
\end{equation}
We then perform the Legendre transform in two steps: first with respect to $%
\phi$ and $G$ only, and then with respect to the rest of the Green
functions. The first (partial) Legendre transform yields

\begin{equation}
\Gamma _\infty [\phi ,G,\{C_r\}]\equiv \Gamma _2[\phi ,G,\{j_n\}]-\sum_{n\ge
3}{\frac 1{n!}}J_n\sum_{P_n}\prod_pC_r
\end{equation}
Here $\Gamma _2$ is the two particle-irreducible (2PI) effective action \cite
{pi2}

\begin{equation}
\Gamma _2[\phi ,G,\{J_n\}]=S[\phi ]+{\frac 12}G^{jk}S,_{jk}-{\frac i2}\ln ~%
{\rm Det}~G+J_{int}[\phi ]+{\frac 12}G^{jk}J_{int,jk}+W_2  \label{gamma2}
\end{equation}
and $W_2$ is the sum of all 2PI vacuum bubbles of a theory whose action is

\begin{equation}
S^{\prime}[\varphi ]={\frac{i}{2}}G^{-1}\varphi^2+S_Q[\varphi ]
\end{equation}

\begin{equation}
S_Q[\varphi ]=S_I[\phi +\varphi ]-S_I[\phi ]-S_I[\phi ],_i\varphi ^i-{\frac 1%
2}S_I[\Phi ],_{ij}\varphi ^i\varphi ^j  \label{eseq}
\end{equation}
where $\varphi $ is the fluctuation field around $\phi $, i.e., $\Phi =\phi
+\varphi $. Decomposing $S_Q$ into source-free and source-dependent parts,
and Taylor expanding with respect to $\varphi $, we may define the
background-field dependent coupling and sources where

\begin{equation}
\sigma_{ni_1...i_n}=\sum_{m\ge n}{\frac{1}{(m-n)!}}
S_{mi_1...i_nj_{n+1}...j_m}\phi^{j_{n+1}}...\phi^{j_m}
\end{equation}

\begin{equation}
\chi_{ni_1...i_n}=\sum_{m\ge n}{\frac{1}{(m-n)!}}
J_{mi_1...i_nj_{n+1}...j_m}\phi^{j_{n+1}}...\phi^{j_m}
\end{equation}
Now, from the properties of the Legendre transformation, we have, for $n>2$,

\begin{equation}
{\frac{\delta W}{\delta J_{n}}}\vert_{J_1,J_2}\equiv {\frac{%
\delta\Gamma_{\infty}}{\delta J_{n}}}\vert_{\phi ,G}
\end{equation}
Computing this second derivative explicitly, we conclude that

\begin{equation}
{\frac{\delta W}{\delta J_{n}}}\vert_{J_1,J_2}\equiv {\frac{1}{n!}}\phi^n +{%
\frac{1}{2~(n-2)!}}G\phi^{n-2}+\sum_{m=3}^n {\frac{\delta\chi_m}{\delta J_n}}
{\frac{\delta W_2}{\delta\chi_{m}}}
\end{equation}
Comparing this equation with

\begin{equation}
{\frac{\delta W}{\delta J_{ni_1...i_n}}}\equiv {\frac{1}{n!}}
\sum_{P_n}\prod_{p}C_r^{j_1...j_r}
\end{equation}
we obtain the identity

\begin{equation}
{\frac{\delta W_2}{\delta \chi _{ni_1...i_n}}}\equiv {\frac 1{n!}}%
\sum_{P_n}^{*}\prod_pC_r^{j_1...j_r}  \label{source2cor}
\end{equation}
where the * above the sum means that clusters containing one element subsets
are deleted. This and

\begin{equation}
\sum_{n\ge 3}{\frac 1{n!}}J_n\sum_{P_n}\prod_pC_r=J_{int}[\phi ]+{\frac 12}%
G^{ij}{\frac{\delta J_{int}[\phi ]}{\delta \phi ^i\delta \phi ^j}}%
+\sum_{n\ge 3}{\frac 1{n!}}\chi _n\sum_{P_n}^{*}\prod_pC_r
\end{equation}
allow us to write 
\begin{eqnarray}
\Gamma _\infty [\phi ,G,\{C_r\}] &\equiv &\ S[\phi ]+({\frac 12})G^{ij}{%
\frac{\delta S[\phi ]}{\delta \phi ^i\delta \phi ^j}}-{\frac i2}\ln ~{\rm Det%
}~G  \label{gamainfty} \\
&&+\{W_2[\phi ,\{\chi _n\}]-\sum_{n\ge 3}{\frac 1{n!}}\chi
_n\sum_{P_n}^{*}\prod_pC_r\}
\end{eqnarray}

This entails an enormous simplification, since it implies that to compute $%
\Gamma_{\infty}$ it is enough to consider $W_2$ as a functional of the $%
\chi_n$, without ever having to decompose these background dependent sources
in terms of the original external sources.

\subsection{Truncation and Slaving: Loop Expansion and Correlation Order}

After obtaining the formal expression for $\Gamma _\infty $, and thereby the
formal hierarchy of Dyson - Schwinger equations, we should proceed with it
much as with the BBGKY hierarchy in statistical mechanics \cite{Huang},
namely, truncate it and close the lower-order equations by constraining the
high order correlation functions to be given (time-oriented) functionals of
the lower correlations. Truncation proceeds by discarding the higher
correlation functions and replacing them by given functionals of the lower
ones, which represent the dynamics in some approximate sense \cite{Akhiezer}%
. The system which results is an open system and the dynamics becomes an
effective dynamics.

It follows from the above that truncations will be generally related to
approximation schemes. In field theory we have several such schemes
available, such as the loop expansion, large $N$ expansions, expansions in
coupling constants, etc. For definiteness, we shall study the case of the
loop expansion, although similar considerations will apply to any of the
other schemes.

Taking then the concrete example of the loop expansion, we observe that the
nonlocal $\chi$ sources enter into $W_2$ in as many nonlinear couplings of
the fluctuation field $\varphi$. Now, $W_2$ is given by a sum of connected
vacuum bubbles, and any such graph satisfies the constraints

\begin{equation}
\sum nV_n=2i
\end{equation}

\begin{equation}
i-\sum V_n=l-1
\end{equation}
where $i,l,V_n$ are the number of internal lines, loops, and vertices with $%
n $ lines, respectively. Therefore,

\begin{equation}
l=1+\sum{\frac{n-2}{2}}V_n
\end{equation}
we conclude that $\chi_n$ only enters the loop expansion of $W_2$ at order $%
n/2$. At any given order $l$, we are effectively setting $\chi_n\equiv 0$, $%
n>2l$. Since $W_2$ is a function of only $\chi_3$ to $\chi_{2l}$, it follows
that the $C_r$'s cannot be all independent. Indeed, the equations relating
sources to Green functions

\begin{equation}
{\frac{\delta W_2}{\delta\chi_{ni_1...i_n}}}\equiv {\frac{1}{n!}}
\sum^*_{P_n}\prod_{p}C_r^{j_1...j_r}
\end{equation}
have now turned, for $n>2l$, into the algebraic constraints

\begin{equation}
\sum ^*_{P_n}\prod_{p}C_r^{j_1...j_r}\equiv 0
\end{equation}
In other words, the constraints which make it possible to invert the
transformation from sources to Green functions allow us to write the higher
Green functions in terms of lower ones. In this way, we see that the loop
expansion is by itself a truncation in the sense above and hence any finite
loop or perturbation theory is intrinsically an effective theory.

Actually, the number of independent Green functions at a given number of
loops is even smaller than $2l$. It follows from the above that $W_2$ must
be linear on $\chi_n$ for $l+2\le n\le 2l$. Therefore the corresponding
derivatives of $W_2$ are given functionals of the $\chi_m$, $m\le l+1$.
Writing the lower sources in terms of the lower order Green functions, again
we find a set of constraints on the Green functions, rather than new
equations defining the relationship of sources to functions. These new
constraints take the form

\begin{equation}
\sum ^*_{P_n}\prod_{p}C_r^{j_1...j_r}=f_n(G,C_3,...C_{l+1})
\end{equation}
for $l+2\le n\le 2l$. In other words, to a given order $l$ in the loop
expansion, only $\phi$, $G$ and $C_r$, $3\le r\le l+1$, enter into $%
\Gamma_{\infty}$ as independent variables. Higher correlations are expressed
as functionals of these by virtue of the constraints implied by the loop
expansion on the functional dependence of $W_2$ on the sources.

However, these constraints are purely algebraic, and therefore do not define
an arrow of time. The dynamics of this lower order functions is unitary.
Irreversibility appears only when one makes a time-oriented ansatz in the
form of the higher correlations, such as the `weakening of correlations'
principle invoked in the truncation of the BBGKY hierarchy \cite{Akhiezer}.
This is done by substituting some of the allowed correlation functions at a
given number of loops $l$, by solutions of the $l$-loop equations of motion.
Observe that even if we use exact solutions, the end result is an
irreversible theory, because the equations themselves are only an
approximation to the true Dyson - Schwinger hierarchy.

To summarize, the truncation of the MEA in a loop expansion scheme proceeds
in two stages. First, for a given accuracy $l$, an $l$- loop effective
action is obtained which depends only on the lowest $l+1$ correlation
functions, say, $\{\phi ,G,C_3,\ldots C_{l+1}\}$. This truncated effective
action generates the $l$-loop equations of motion for these correlation
functions. In the second stage, these equations of motion are solved (with
causal boundary conditions) for some of the correlation functions, say $%
\{C_k,...C_{l+1}\}$, and the result is substituted into the $l$ loop
effective action. (We say that $\{C_k,...C_{l+1}\}$ have been {\it slaved}
to $\{\phi ,G,C_3,...C_{k-1}\}$) The resulting truncated effective action is
generally complex and the mean field equations of motion it generates will
come out to be dissipative, which indicates that the effective dynamics is
stochastic.

\subsection{Example: the three-loop 2PI EA}

We shall conclude this paper by explicitly computing the 2PI CTP EA for a $%
\lambda \phi ^4$ self interacting scalar field theory, out of the
corresponding MEA. We carry out our analysis at three loops order, this
being the lowest order at which the dynamics of the correlations is
nontrivial, in the absence of a symmetry breaking background field \cite
{CH88}.

To this accuracy, we have room for four nonlocal sources besides the mean
field and the two point correlations, namely $\chi _3$, $\chi _4$,$\chi _5$
, and $\chi _6$. However, the last two enter linearly in the generating
functional. Thus the three- loop effective action only depends nontrivially
on the mean field and the two, three and four point correlations. By
symmetry, there must be a solution where the mean field and the three point
function remain identically zero, which we shall assume.

Our first step is to compute Eq. (\ref{gamainfty}), which now reads

\begin{eqnarray}
\Gamma _4[G,C_4] &\equiv &\ ({\frac{-1}2})c_{ij}\left( -\Box +M^2\right)
G^{ij}-{\frac i2}\ln ~{\rm Det}~G  \label{twopi} \\
&&\ \ \ +W_2[\phi ,\{\chi _n\}]-{\frac 1{24}}\chi _{4ijkl}\left[
C_4^{ijkl}+G^{ij}G^{kl}+G^{ik}G^{jl}+G^{il}G^{jk}\right]  \nonumber
\end{eqnarray}
where $W_2$ denotes the sum of 2PI vacuum bubbles of a quantum field theory
with quartic self interaction and a coupling constant $\lambda -\chi _4$
(see eqs. (\ref{gamma2}) and (\ref{eseq})) up to three loops

\begin{eqnarray}
W_2 &=&\left( \frac{-1}8\right) \left( \lambda -\chi _4\right)
_{ijkl}G^{ij}G^{kl} \\
&&+\left( \frac i{48}\right) \left( \lambda -\chi _4\right) _{ijkl}\left(
\lambda -\chi _4\right) _{pqrs}G^{ip}G^{jq}G^{kr}G^{ls}  \nonumber
\end{eqnarray}

Eq. (\ref{source2cor}) yields

\begin{equation}
C_4^{ijkl}=-i\left( \lambda -\chi _4\right) _{pqrs}G^{ip}G^{jq}G^{kr}G^{ls}
\end{equation}

Inverting and substituting back in Eq. (\ref{twopi}), we obtain

\begin{eqnarray}
\Gamma _4[G,C_4] &\equiv &\ ({\frac{-1}2})c_{ij}\left( -\Box +M^2\right)
G^{ij}-{\frac i2}\ln ~{\rm Det}~G  \label{threeloops} \\
&&-\left( \frac 18\right) \lambda _{ijkl}G^{ij}G^{kl}-\left( \frac 1{24}%
\right) \lambda _{ijkl}C_4^{ijkl}  \nonumber \\
&&+\left( \frac i{48}\right) C_4^{ijkl}\left[
G_{ip}^{-1}G_{jq}^{-1}G_{kr}^{-1}G_{ls}^{-1}\right] C_4^{pqrs}  \nonumber
\end{eqnarray}
This functional generates the self consistent, time reversal invariant
dynamics of the two and four particle Green functions to three loop
accuracy. To reduce it further to the dynamics of the two point functions 
{\it alone}, we must {\it slave} the four point functions. Consider the
three loops equation of motion for $C_4$

\begin{equation}
\left[ G_{ip}^{-1}G_{jq}^{-1}G_{kr}^{-1}G_{ls}^{-1}\right]
C_4^{pqrs}=-i\lambda _{ijkl}
\end{equation}

Solving for this equation with causal boundary conditions yields

\begin{equation}
C_4^{ijkl}=-i\lambda _{pqrs}G^{ip}G^{jq}G^{kr}G^{ls}
\end{equation}

(in other words, $\chi _4=0$) and substituting back in Eq. (\ref{threeloops}%
) we obtain

\begin{eqnarray}
\Gamma [G] &\equiv &\ ({\frac{-1}2})c_{ij}\left( -\Box +M^2\right) G^{ij}-{%
\frac i2}\ln ~{\rm Det}~G  \label{redloops} \\
&&\ \ \ -\left( \frac 18\right) \lambda _{ijkl}G^{ij}G^{kl}  \nonumber \\
&&\ \ +\left( \frac i{48}\right) \lambda
_{ijkl}G^{ip}G^{jq}G^{kr}G^{ls}\lambda _{pqrs}  \nonumber
\end{eqnarray}
which is seen to be equivalent to Eq. (\ref{gamma2pib}). This effective
action leads to a dissipative and, as we have seen, also stochastic
dynamics, which results from the slaving of the four point functions.

\section{Discussions}

In this paper we have introduced a new object, the stochastic propagator $%
{\bf G}$, whose expectation value reproduces the usual propagators, but
whose fluctuations are designed to account for the quantum fluctuations in
the binary product of fields. We have introduced the dynamical equation for $%
{\bf G}$ which takes the form of an explicitly stochastic Schwinger - Dyson
equation, and showed that in the kinetic limit, both the fluctuations in $%
{\bf G}$ become the classical fluctuations in the one particle distribution
function, and the dynamic equation for ${\bf G}$'s Wigner transform becomes
the Boltzmann - Langevin equation. Each of these results has interest of its
own. A priori, there is no simple reason why the fluctuations derived from
quantum field theory should have a physical meaning corresponding to a
phenomenological entropy flux and Einstein's relation.

The notion that Green functions (and indeed, higher correlations as well)
may or even ought to be seen as possessing fluctuating characters (when
placed in the larger context of the whole hierarchy) with clearly
discernable physical meanings is likely to have an impact on the way we
perceive the statistical properties of field theory. For example, we are
used to fixing the ambiguities of renormalization theory by demanding
certain Green functions to take on given values under certain conditions
(conditions which should resemble the physical situation of interest as much
as possible, as discussed by O'Connor and Stephens \cite{OCSte}). If the
Green functions themselves are to be regarded as fluctuating, then the same
ought to hold for the renormalized coupling constants defined from them, and
to the renormalization group (RG) equations describing their scale
dependence.

While the application of renormalization group methods to stochastic
equations is presented in well-known monographs\cite{Zinn}, our proposal
here goes beyond these results in at least two ways. First, in our approach
the noise is not put in by hand or brought in from outside (e.g., the
environment of an open system), as in the usual Langevin equation approach,
but it follows from the (quantum) dynamics of the system itself. Actually,
the possibility of learning about the system from the noise properties
(whether it is white or coloured, additive or multiplicative, etc.) -- {\it %
unraveling} the noise, or treating noise {\it creatively}-- is a subtext in
our program. Second, our result suggests that stochasticity may, or should,
reside beyond the level of equations of motion, and appear at the level of
the RG equations, as they describe the running of quantities which are
themselves fluctuating.

Indeed, the possibility of a nondeterministic renormalization group flow is
even clearer if we think of the RG as encoding the process of eliminating
irrelevant degrees of freedom from our description of a system \cite{Ma} .
These elimination processes lead as a rule to dissipation and noise, the
noise and dissipation in the influence action and the CTP-effective action
are but a particular case. If the need for such an enlarged RG has not been
felt so far, the groundbreaking work on the dynamical RG by Ma, Mazenko,
Hohenberg, Halperin, and many significant others notwithstanding, it is
probably due to the fact that the bulk of RG research has been focused on
equilibrium, stationary properties rather than far- from- equilibrium
dynamics\cite{dcf}.\\

{\bf Acknowledgements} EC is supported in part by CONICET, UBA and
Fundaci\'on Antorchas. BLH is supported in part by NSF grant PHY98-00967 and
their collaboration is supported in part by NSF grant INT95-09847. A
preliminary version of this work was presented at the RG2000 meeting in
Taxco (Mexico), Jan. 1999. We wish to thank the organizers for this
excellent meeting and their lavish hospitality, especially Denjoe O 'Connor
and Chris Stephens, with whom we enjoyed many close discussions over the
years. We also appreciate the interest expressed by David Huse and exchanges
with Michael Fisher and Jean Zinn - Justin during the meeting on noise in
nonequilibrium renormalization group theory.

\section{Appendix: closed time path conventions}

The closed time path (CTP)\ or Schwinger - Keldysh technique \cite{ctp} is a
bookeeping device to generate diagrammatic expansions for true expectation
values (as opposed to IN - OUT matrix elements) of certain quantum
operators. The basic idea is that any expectation value of the form

\begin{equation}
\left\langle IN\left| \tilde T\left[ \phi \left( x_1\right) ...\phi \left(
x_n\right) \right] T\left[ \phi \left( x_{n+1}\right) ...\phi _m\right]
\right| IN\right\rangle  \label{app1}
\end{equation}

where $\left| IN\right\rangle $ is a suitable initial quantum state, $x_1$
to $x_m$ are space time points, $\phi $ is the field operator, $T$ stands
for time ordering and $\tilde T$ for anti time ordering, may be thought of
as a path ordered expectation value on a closed time - path ranging from $%
t=-\infty $ to $\infty $ and back. These path ordered products are generated
by path integrals of the form

\begin{equation}
\int D\phi ^1D\phi ^2\;\left[ \phi ^2\left( x_1\right) ...\phi ^2\left(
x_n\right) \phi ^1\left( x_{n+1}\right) ...\phi _m^1\right] e^{i\left[
S\left( \phi ^1\right) -S^{*}\left( \phi ^2\right) \right] }  \label{app2}
\end{equation}

where $\phi ^1$ is a field configuration in the forward leg of the path, and 
$\phi ^2$ likewise on the return leg. These configurations match each other
on a spacelike surface at the distant future. The boundary conditions at the
distant past depend on the initial state $\left| IN\right\rangle $; for
example, if this is a vacuum, then we add a negative imaginary part to the
mass. We shall not discuss these boundary conditions further, except to note
that we assume the validity of Wick's theorem (see \cite{CH88}).

In general we shall use a latin index $a$, $b$,.... taking values $1$ or $2$
to denote the CTP branches. Where the space time position is not specified,
it must be assumed that it has been subsummed within the CTP upper index.
Also we shall refer to the expression $S\left( \phi ^a\right) =S\left( \phi
^1\right) -S^{*}\left( \phi ^2\right) $ as the CTP action. We allways use
the Einstein sum convention, and if not explicit, integration over space
time must be understood as well.

It is convenient to introduce a CTP metric tensor $c_{ab}=diag(1,-1)$ to
keep track of sign inversions. Thus $c_{ab}J^a\phi ^b=J^1\phi ^1-J^2\phi ^2$%
. In general, we write an expression like this as $J_a\phi ^a$, where $%
J_a=c_{ab}J^b;$ the index $a$ has been lowered by means of the metric
tensor. The opposite operation of raising an index is accomplished with the
inverse metric tensor $c^{ab}=\left( c^{-1}\right) ^{ab}=diag(1,-1)$. Thus $%
J^a=c^{ab}J_b$ .

\end{document}